\documentclass[conference,a4paper]{IEEEtran}

\usepackage[utf8]{inputenc} 
\usepackage[T1]{fontenc}
\usepackage{url}
\usepackage{hyperref}
\usepackage{ifthen}
\usepackage{cite}
\usepackage{stfloats}
\usepackage[cmex10]{amsmath} 
\usepackage[pdftex]{graphicx}
\ifCLASSOPTIONcompsoc
  \usepackage[caption=false,font=normalsize,labelfont=sf,textfont=sf]{subfig}
\else
\usepackage[caption=false,font=footnotesize]{subfig}
\fi
\usepackage{enumitem}
\DeclareGraphicsExtensions{.pdf}

\usepackage{amssymb}
\newtheorem{theorem}{Theorem}
\newtheorem{corollary}{Corollary}
\newtheorem{lemma}{Lemma}

\begin{document}
\title{A Simplified Coding Scheme for the Broadcast Channel With Complementary Receiver Side Information Under Individual Secrecy Constraints} 

\author{%
   \IEEEauthorblockN{Jin Yeong Tan, Lawrence Ong, and Behzad Asadi}
   \IEEEauthorblockA{School of Electrical Engineering and Computing, The University of Newcastle, Newcastle, Australia\\
                     Email: jinyeong.tan@uon.edu.au, lawrence.ong@newcastle.edu.au, behzad.asadi@uon.edu.au}
 }

\maketitle

\begin{abstract}
	This paper simplifies an existing coding scheme for the two-receiver discrete memoryless broadcast channel with complementary receiver side information where there is a passive eavesdropper and individual secrecy is required. The existing coding scheme is simplified in two steps by replacing Wyner secrecy coding with Carleial-Hellman secrecy coding. The resulting simplified scheme is free from redundant message splits and random components. Not least, the simplified scheme retains the existing achievable individual secrecy rate region. Finally, its construction simplicity helps us gain additional insight on the integration of secrecy techniques into error-correcting coding schemes. 
\end{abstract}



\section{Introduction}

\subsection{Background}

The open nature of wireless communication channels makes transmitted signals susceptible to passive eavesdroppers present in the communication range. As a result, the problem of secure communication channel has always been of utmost priority. As such, various security techniques have been developed for application at different layers of the protocol stack such as cryptography and fequency hopping. Recently, information theoretic secrecy has gained more attention since it has significant advantages by not considering assumptions of limited computational power of the eavesdropper.

Information theoretic secrecy in the presence of eavesdroppers was first applied in point-to-point communication systems and the earliest study dates back to Shannon's paper in 1949 \cite{cref8}. In the paper, Shannon proposed a secure communication strategy by having the transmitter and legitimate receiver share a secret key, which is unknown to the eavesdropper. This secret key approach (also known as one-time pad) which achieves perfect secrecy, i.e., zero information leakage, however, requires the secret key to be at least the size of the message and be independent of the message. 

The statistical difference in channel noise has also been exploited as a means to ensure information theoretic secrecy. This is demonstrated by Wyner who introduced the wiretap channel in which the signal transmitted by a sender is received by a legitimate receiver and an eavesdropper via noisy channels \cite{cref4}. The wiretap coding scheme invovles multicoding and randomized encoding. The principle behind this coding scheme is to provide sufficient randomness to the codeword in order to protect the message from the eavesdropper. However, secret communication is viable through this scheme if the channel to the eavesdropper is weaker than the channel to the legitimate receiver. 

The common approaches of information theoretic secrecy discussed above were even applied in broadcast channels. Schaefer and Khisti implemented the usage of secret keys to securely transmit a common message to two legitimate receivers while keeping the message ignorant from an eavesdropper \cite{Schaefer_Khisti14}. On the other hand, Csisz$\acute{\text{a}}$r and K$\ddot{\text{o}}$rner \cite{iref1} as well as Chia and El Gamal \cite{iref2} extended Wyner wiretap coding to the general discrete memoryless broadcast channels. The cases in \cite{iref1}, \cite{iref2} study broadcast channels with different number of receivers but both involve securing a single private message. A combinational approach was even demonstrated by Chen, Koyluoglu and Sezgin \cite{cref2} in which both the one-time pad approach and Wyner serecy coding are combined into the superposition-Marton coding scheme \cite{cref6,cref5} to ensure secure broadcasting in the case of the two-receiver discrete memoryless broadcast channel with complementary receiver side information and with a passive eavesdropper.



Over the course of the aforementioned studies, there is no single unified strategy while integrating secrecy techniques into existing error-correcting coding schemes (without secrecy constraints) to ensure information theoretic secrecy. This motivates us to investigate if there exists a more intuitive guide on applying secrecy techniques to existing error-correcting coding schemes when secrecy is concerned. To achieve this goal, we continue from the study of Chen et al. \cite{cref2} which focus on the two-receiver discrete memoryless broadcast channel with complementary receiver side information and with an eavesdropper under individual secrecy constraints. Our choice of study follows upon the fact that Chen-Koyluoglu-Sezgin secrecy coding scheme is more comprehensive since it combines two secrecy techniques into the popular superposition-Marton error-correcting coding scheme. In short, this paper aims to obtain an intuitive understanding on the integration of individual secrecy techniques to existing channel coding schemes.

\subsection{Contributions}

We notice that Carleial-Hellman secrecy coding \cite{cref3}, which allows each message to act as a random component in ensuring the individual secrecy of each another, can be used to simplify the stucture of Chen-Koyluoglu-Sezgin secrecy coding scheme for the two-receiver discrete memoryless broadcast channel with complementary receiver side information and with a passive eavesdropper. This leads us to perform a two-step scheme simplification which involves the gradual replacement of Wyner secrecy coding with Carleial-Hellman secrecy coding. The final simplified coding scheme essentially allows us to identify the redundant components in the existing coding scheme. Although the proposed simplified coding scheme is shown to achieve the same rate region as the existing coding scheme, its construction simplicity provides a clear view on the integration of secrecy techniques into existing error-correcting coding schemes and enriches our present understanding on coding principles for individual secrecy. This understanding also acts as a basis while devising a coding scheme for secure broadcasting in the two-receiver discrete memoryless broadcast channel with one-sided receiver side information in our parallel work \cite{Me2}.

\subsection{Paper Organization}

The entire paper will be organized as follows. Section II will focus on the system model. Section III will provide an overview of the related works that motivate this paper. Section IV will provide the main results on the coding scheme simplification. Lastly, section V will provide a brief discussion and conclude the paper. 

\section{System Model}

In this paper, we will denote random variables by uppercase letters, their corresponding realizations by lowercase letters and their corresponding sets by calligraphic letters. A $(j-i+1)$-sequence of random variables will be denoted by $X_i^j=(X_i,…,X_j)$ for $1\leq i\leq j$. Whenever $i=1$, the subscript will be dropped, resulting in $X^j=(X_1,…,X_j)$. $\mathbb{R}^d$ represents the $d$-dimensional real Euclidean space and $\mathbb{R}_+^d$ represents the $d$-dimensional non-negative real Euclidean space. $\mathcal{R}$ will be used to represent a subset of $\mathbb{R}^d$, and the convex hull of $\mathcal{R}$ will be denoted as co$(\mathcal{R})$. Meanwhile, $[a:b]$ refers to a set of natural numbers between and including $a$ and $b$, for $a\leq b$. Lastly, the operator $\times$ denotes the Cartesian product.

\begin{figure}[t]
	\centering
	\includegraphics[width=2.5in,height=1in,keepaspectratio]{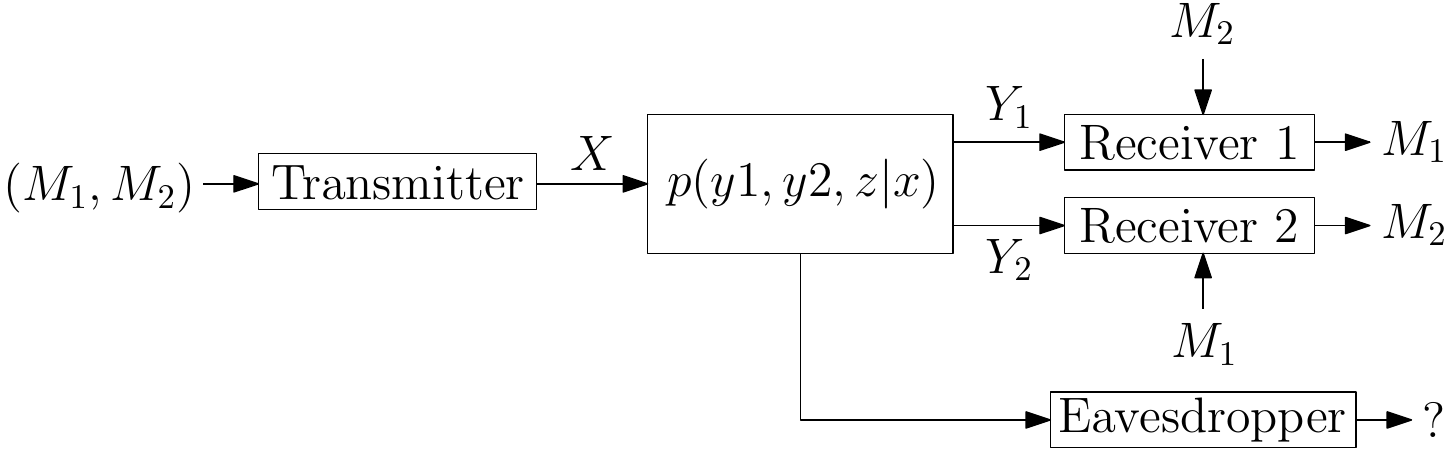}
	\caption{The two-receiver broadcast channel with an eavesdropper.}
	\label{fig:1}
\end{figure}
The focus of this paper will be on the scheme simplification for the two-receiver discrete memoryless broadcast channel with complementary receiver side information and with a passive eavesdropper (hereafter referred as the broadcast channel with complementary side information). The system model for this case is illustrated in Fig.~\ref{fig:1}. In this model, we define $(M_1, M_2)$ as the source messages, $M_i$ as the message requested by legitimate receiver $i$ and $M_{\hat{\imath}}$, $\hat{\imath}\triangleq (i\bmod 2)+1$ as the message known a priori (i.e., receiver side information), for all $i\in\{1,2\}$. Let $X$ denote the channel input from the sender, while $Y_i$ and $Z$ denote the channel output to receiver $i$ and the eavesdropper respectively. In $n$ channel uses,  $X^n$ represents the transmitted codeword, $Y_i^n$ represents the signal received by legitimate receiver $i$ and $Z^n$ represents the signal received by the eavesdropper. The memoryless (and without feedback) nature of the channel also implies that
\begin{align}
&p(y_1^n,y_2^n,z^n|x^n)=\prod_{i=1}^n p(y_{1i},y_{2i},z_i|x_i)
\end{align} 

In this case, the transmitter will be sending messages $M_{1}$ and $M_{2}$ to legitimate receiver 1 and 2, respectively through the channel $p(y_{1},y_{2},z|x)$. Both legitimate receivers have complementary receiver side information to aid them in decoding the transmitted messages. In other words, receiver 1 will be having $M_2$ as side information, whereas receiver 2 will be having $M_1$ as side information.

\textit{Definition 1:} A $(2^{nR_1},2^{nR_2},n)$ secrecy code for the two-receiver discrete memoryless broadcast channel with complementary side information consists of
\begin{itemize}
	\item two message sets, where $\mathcal{M}_1=[1:2^{nR_1}]$ and $\mathcal{M}_2=[1:2^{nR_2}]$;
	\item an encoding function, $f:\mathcal{M}_1\times\mathcal{M}_2\rightarrow\mathcal{X}^n$, such that $X^n=f(M_1,M_2)$; and
	\item two decoding functions, $g_i:\mathcal{Y}_i^n\times\mathcal{M}_{\hat{\imath}}\rightarrow\mathcal{M}_i$, such that $\hat{M}_i=g_i(Y_i^n, M_{\hat{\imath}})$, where $\hat{\imath}\triangleq (i\bmod 2)+1$, for all $i\in\{1,2\}$.
\end{itemize}

Both messages, $M_1$ and $M_2$ are assumed to be uniformly distributed over their respective message set. Hence, we have $R_i=\frac{1}{n}H(M_i)$, for all $i\in\{1,2\}$. Meanwhile the individual information leakage rate associated with the $(2^{nR_1},2^{nR_2},n)$ secrecy code is defined as $R_{\text{L},i}^{(n)}=\frac{1}{n}I(M_i;Z^n)$, for all $i\in\{1,2\}$.

The probability of error for the secrecy code at each receiver $i$ is defined as $P_{\text{e},i}^{(n)}=P\{\hat{M}_i\ne M_i\}$, for $i\in\{1,2\}$. A rate pair $(R_1,R_2)$ is said to be achievable if there exist a sequence of $(2^{nR_1},2^{nR_2},n)$ codes such that 
\begin{align}
&P_{\text{e},i}^{(n)}\leq \epsilon_n\text{, for all }i\in\{1,2\}\\			
&R_{\text{L},i}^{(n)}\leq \tau_n\text{, for all }i\in\{1,2\}\\		
& \lim\limits_{n\rightarrow\infty}\epsilon_n=0\text{ and }\lim\limits_{n\rightarrow\infty}\tau_n=0 			
\end{align}

\section{Related Works}

\begin{figure*}[t]
	\centerline{\subfloat[Rate splitting]{\includegraphics[width=2in,height=0.6in,keepaspectratio]{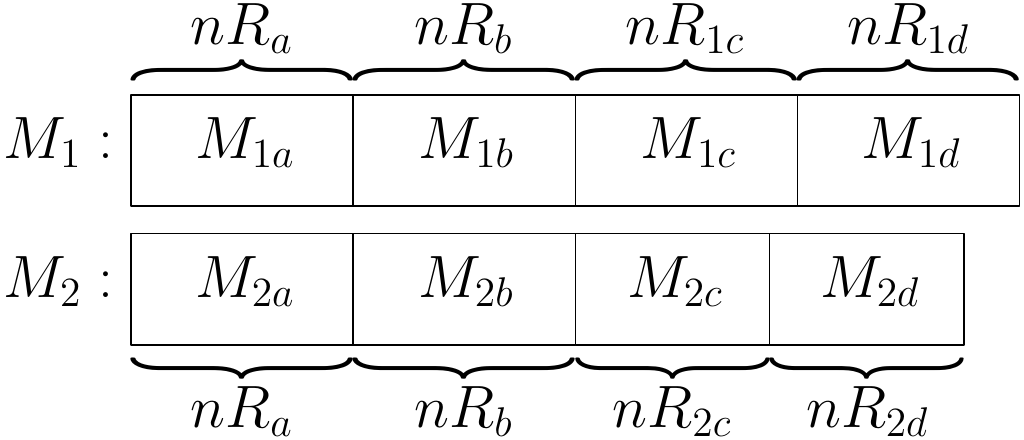}
			\label{fig:3a}}
		\hfil
	\subfloat[Encoding]{\includegraphics[width=0.45\textwidth]{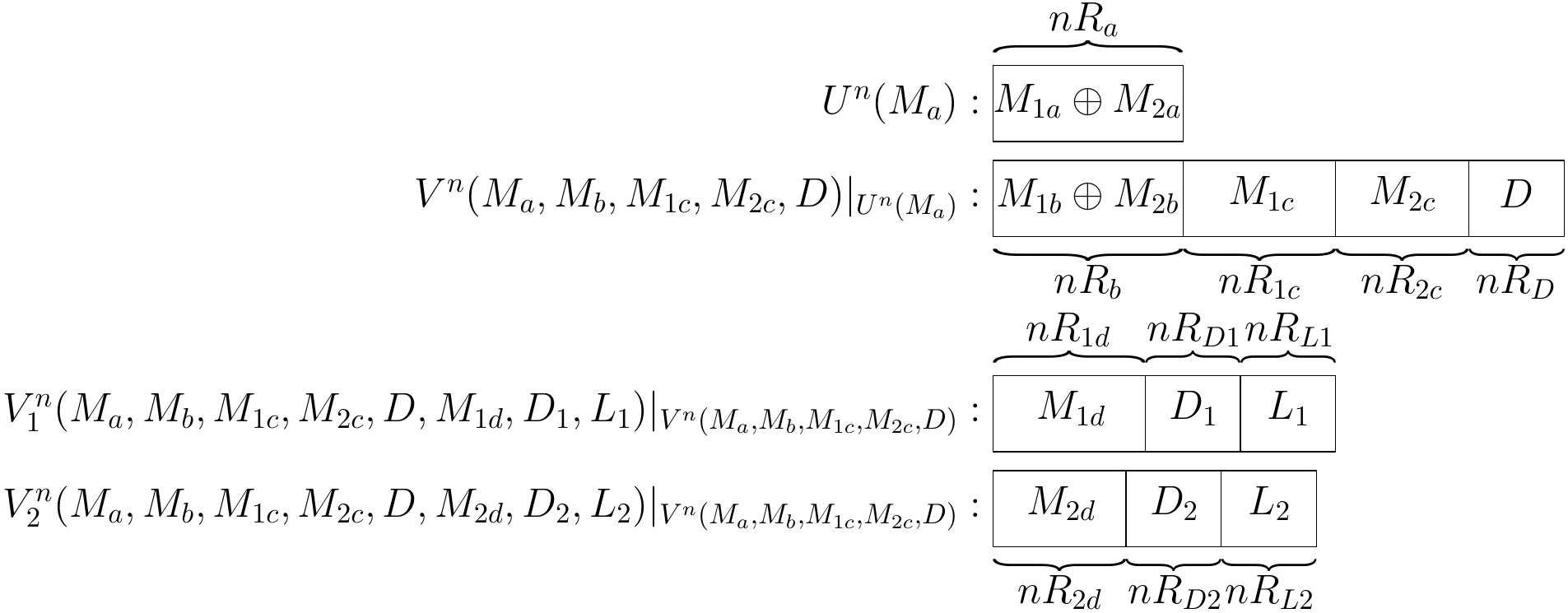}
			\label{fig:3b}}}
	\caption{An existing coding scheme for the two-receiver broadcast channel with complementary receiver side information and with an eavesdropper.}
	\label{fig:3}
\end{figure*}

The problem in this paper is motivated by the discussions of Chen et al. \cite{cref2}. Chen et al. proposed a coding scheme for the broadcast channel with complementary side information under the context of individual secrecy. As illustrated in Fig.~\ref{fig:3a}, this coding scheme involves splitting $M_i$, for all $i\in\{1,2\}$, into four independent message segments, namely $M_{ia}$ at rate $R_{a}$, $M_{ib}$ at rate $R_{b}$, $M_{ic}$ at rate $R_{ic}$ and $M_{id}$ at rate $R_{id}$, i.e., $M_i=(M_{ia},M_{ib},M_{ic},M_{id})$ where $M_{ia}\in [1:2^{nR_{a}}]$, $M_{ib}\in [1:2^{nR_{b}}]$, $M_{ic}\in [1:2^{nR_{ic}}]$, $M_{id}\in [1:2^{nR_{id}}]$ and $R_i=R_{a}+R_{b}+R_{ic}+R_{id}$. These message segments then make up the coding scheme in Fig.~\ref{fig:3b} which comprises of four layers: cloud center codeword $U^n$, common satellite codeword $V^n$ as well as private satellite codewords $V_1^n$ and $V_2^n$. 

In this scheme, the cloud center codeword $U^n$ comprises a one-time pad signal \cite{cref8}. This secret key $M_{1a}\oplus M_{2a}$ is formed by XOR-ing the first segment of each message. The utilization of a one-time pad signal is possible since the legitimate receivers can utilize their receiver side information to unveil the original message. 

A main contribution in the coding scheme proposed by Chen et al. \cite{cref2} is a common satellite codeword $V^n$ formed and combined with $U^n$ using superposition coding \cite{cref6}. In the $V^n$ codeword layer, Wyner secrecy coding \cite{cref4} is implemented to protect the message segments $M_{1c}$ and $M_{2c}$ using a randomly generated $D$. However, apart from the regular randomness $D$, the authors presented an interesting idea of providing additional randomness through another one time-pad signal $M_{1b}\oplus M_{2b}$. 

Meanwhile, the private satellite codewords $V_1^n$ and $V_2^n$, which are responsible of carrying the remaining message segments $M_{1d}$ and $M_{2d}$ to their respective legitimate receivers, are also secured with Wyner secrecy coding \cite{cref4} using randomly generated $D_1$ and $D_2$ respectively. The private satellite codewords $V_1^n$ and $V_2^n$ are first combined under Marton coding in which the components $L_1$ and $L_2$ are present to ensure the existence of at least one jointly typical sequence pair between $V_1^n$ and $V_2^n$ \cite{cref5}. Next, the $V_1^n$ and $V_2^n$ codeword layers are combined with the upper $U^n$ and $V^n$ codeword layers using superposition coding, forming a superposition-Marton coding framework to achieve an improved rate region. This coding scheme has the achievability conditions listed in Theorem~\ref{theorem2}. After the Fourier-Motzkin procedure \cite[Appendix D]{cref6}, the achievable individual secrecy rate region $\mathcal{R}^{\text{old}}$ is obtained as in Corollary~\ref{corollary 2}. 

\begin{theorem}[Chen et al.~\cite{cref2}]\label{theorem2}
	The following individual secrecy rate region is achievable for the two-receiver discrete memoryless broadcast channel with complementary receiver side information and with a passive eavesdropper:
	\begin{equation}\label{eq1}
	\left\{
	\begin{split}
	(R_1,R_2)\\ \in \mathbb{R}_+^2
	\end{split}
	\left\vert
	\begin{split}
	&R_1=R_a+R_b+R_{1c}+R_{1d},\\
	&R_2=R_a+R_b+R_{2c}+R_{2d},\\
	&R_{L1}+R_{L2}>I(V_1;V_2|V),\\
	&R_1+R_D+R_{D1}+R_{L1}<I(U,V,V_1;Y_1),\\
	&R_1-R_a+R_D+R_{D1}+R_{L1}<I(V,V_1;Y_1|U),\\
	&R_{1d}+R_{D1}+R_{L1}<I(V_1;Y_1|V),\\
	&R_2+R_D+R_{D2}+R_{L2}<I(U,V,V_2;Y_2),\\
	&R_2-R_a+R_D+R_{D2}+R_{L2}<I(V,V_2;Y_2|U),\\
	&R_{2d}+R_{D2}+R_{L2}<I(V_2;Y_2|V),\\
	&R_{D1}+R_{L1}\geq I(V_1;Z|V),\\
	&R_{D2}+R_{L2}\geq I(V_2;Z|V),\\
	&R_b+R_D\overset{\text{(a)}}{\geq}I(V;Z|U),\\
	&R_{L1}+R_{L2}\leq I(V_1;Z|V)+I(V_2;Z|V)\\&-I(V_1,V_2;Z|V),\\
	&R_1\geq 0,R_a\geq 0,R_b\geq 0,R_{1c}\geq 0,\\
	&R_{1d}\geq 0,R_2\geq 0,R_{2c}\geq 0,R_{2d}\geq 0,\\
	&R_D\geq 0,R_{D1}\geq 0,R_{D2}\geq 0,R_{L1}\geq 0,R_{L2}\geq 0\\
	&\text{over all } p(u)p(v|u)p(v_1,v_2|v)p(x|v_1,v_2)
	\end{split}
	\right.
	\right\}
	\end{equation}
\end {theorem}
	
\begin{corollary}[{\cite[Theorem 9]{cref2}}]\label{corollary 2}
	The achievable individual secrecy rate region in Theorem~\ref{theorem2} can be written as follows:
	\begin{equation}\label{eq2}
	\mathcal{R}^{\text{old}} \triangleq
	\left\{
	\begin{split}
	(R_1,R_2)\\ \in \mathbb{R}_+^2
	\end{split}
	\left\vert
	\begin{split}
    &R_1<I(U;Y_1)+I(V,V_1;Y_1|U)\\&-I(V,V_1;Z|U)+I(V;Z|U),\\
    &R_1-R_2<I(V,V_1;Y_1|U)-I(V,V_1;Z|U),\\
    &R_2<I(U;Y_2)+I(V,V_2;Y_2|U)\\&-I(V,V_2;Z|U)+I(V;Z|U),\\
    &R_2-R_1<I(V,V_2;Y_2|U)-I(V,V_2;Z|U)\\
	&\text{over all }p(u)p(v|u)p(v_1,v_2|v)p(x|v_1,v_2)\\
	&\text{ subject to }I(V_1;V_2|V)\overset{\text{(a)}}{<}I(V_1;Z|V)\\&+I(V_2;Z|V)-I(V_1,V_2;Z|V),\\
	&I(V,V_i;Y_i|U)\overset{\text{(b)}}{>}I(V,V_i;Z|U)\text{ and }\\
	&I(V_i;Y_i|V)\overset{\text{(c)}}{>}I(V_i;Z|V)\text{, for all }i\in\{1,2\}. 
	\end{split}
	\right.
	\right\}
	\end{equation}
\end{corollary}

Note that for the superscript of $\mathcal{R}$, we use the shorthand notation 'old' to denote an existing region and 'new' to denote a new region. We will also use (\textit{xx}.\textit{yy}) to denote the \textit{yy} constraint in equation \textit{xx}.

\section{Proposed Simplification for the Two-Receiver Broadcast Channel with Complementary Side Information}
The coding scheme simplification for the two-receiver broadcast channel with complementary side information was carried out in two steps.

\begin{figure*}[t]
	\centerline{\subfloat[Rate splitting]{\includegraphics[width=2in,height=0.6in,keepaspectratio]{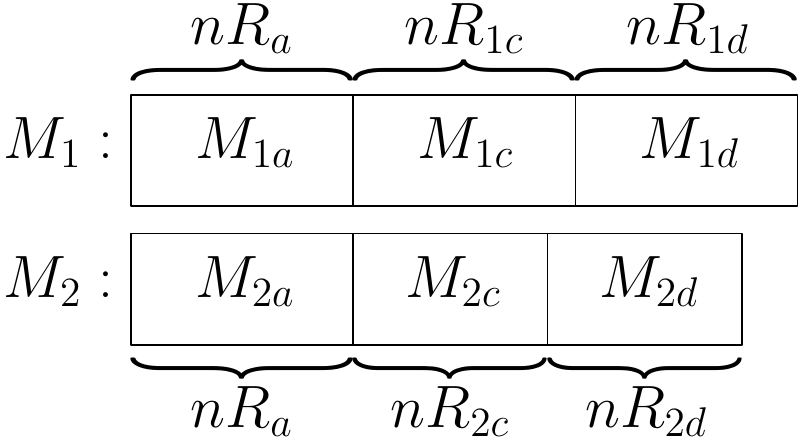}
			\label{fig:6a}}
		\hfil
	\subfloat[Encoding]{\includegraphics[width=0.35\textwidth]{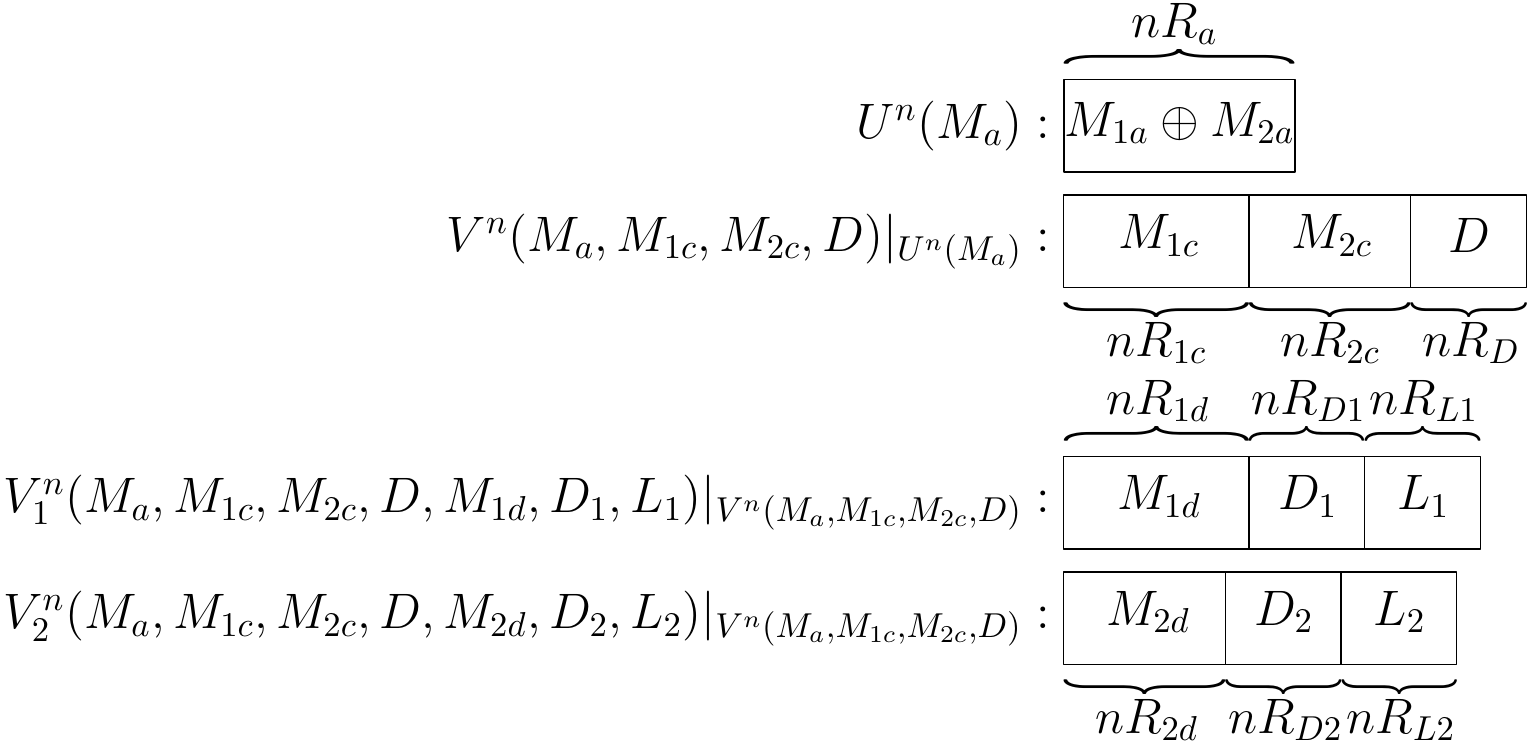}
			\label{fig:6b}}}
		\caption{Simplified coding scheme 1.}
		\label{fig:6}
	\end{figure*}

\subsection{Simplification Step 1}

For the first simplification step, we performed rate splitting to break $M_i$, for all $i\in\{1,2\}$, into three indepedent message segments as shown in Fig.~\ref{fig:6a}. Compared to rate splitting for the existing coding scheme in Fig.~\ref{fig:3a}, our approach requires one less message split for $M_i$, i.e., $M_{ib}=\phi$. In line with the existing coding scheme, our proposed simplified coding scheme 1 is then constructed from a combination of a one-time pad signal \cite{cref8}, superposition coding \cite{cref6}, Wyner secrecy coding \cite{cref4} and Marton coding \cite{cref5} as illustrated in Fig.~\ref{fig:6b}.

Our simplified coding scheme 1 differs from the existing coding scheme \cite{cref2} in Fig.~\ref{fig:3} as we bring in Carleial-Hellman secrecy coding \cite{cref3} to partially replace Wyner secrecy coding \cite{cref4} in the common satellite codeword  $V^n$. Through this approach, the message segments $M_{1c}$ and $M_{2c}$ are now responsible of ensuring the individual secrecy of each another with the aid of an additional randomness $D$. 

The partial replacement of Wyner secrecy coding \cite{cref4} with Carleial-Hellman secrecy coding \cite{cref3} changes the individual secrecy analysis. Hence, it is insufficient to directly set $R_b=0$ in order to reflect $M_{ib}=\phi$. A modified decoding strategy is required, resulting in Lemma~\ref{lemma 1}. Using $\mathcal{R}^{\text{old'}}$ in Lemma~\ref{lemma 1}, we then obtain the achievable individual secrecy rate region of the simplified coding scheme 1 as presented in Theorem~\ref{theorem5}. 

\begin{lemma}\label{lemma 1}
	The existing coding scheme by Chen et al. \cite{cref2} for the two-receiver discrete memoryless broadcast channel with complementary receiver side information and with a passive eavesdropper also achieves an alternative individual secrecy rate region $\mathcal{R}^{\text{old'}}$ as stated below:
	\begin{equation}\label{eq3}
	\left\{
	\begin{split}
	(R_1,R_2)\\ \in \mathbb{R}_+^2
	\end{split}
	\left\vert
	\begin{split}
	&R_1=R_a+R_b+R_{1c}+R_{1d},\\
	&R_2=R_a+R_b+R_{2c}+R_{2d},\\
	&R_{L1}+R_{L2}>I(V_1;V_2|V),\\
	&R_1+R_D+R_{D1}+R_{L1}<I(U,V,V_1;Y_1),\\
	&R_1-R_a+R_D+R_{D1}+R_{L1}<I(V,V_1;Y_1|U),\\
	&R_{1d}+R_{D1}+R_{L1}<I(V_1;Y_1|V),\\
	&R_2+R_D+R_{D2}+R_{L2}<I(U,V,V_2;Y_2),\\
	&R_2-R_a+R_D+R_{D2}+R_{L2}<I(V,V_2;Y_2|U),\\
	&R_{2d}+R_{D2}+R_{L2}<I(V_2;Y_2|V),\\
	&R_{D1}+R_{L1}\geq I(V_1;Z|V),\\
	&R_{D2}+R_{L2}\geq I(V_2;Z|V),\\
	&R_b+R_{1c}+R_D\overset{\text{(a)}}{\geq}I(V;Z|U),\\
	&R_b+R_{2c}+R_D\overset{\text{(b)}}{\geq}I(V;Z|U),\\
	&R_{L1}+R_{L2}\leq I(V_1;Z|V)+I(V_2;Z|V)\\&-I(V_1,V_2;Z|V),\\
	&R_1\geq 0,R_a\geq 0,R_b\geq 0,R_{1c}\geq 0,\\
	&R_{1d}\geq 0,R_2\geq 0,R_{2c}\geq 0,R_{2d}\geq 0,\\
	&R_D\geq 0,R_{D1}\geq 0,R_{D2}\geq 0,R_{L1}\geq 0,R_{L2}\geq 0\\
	&\text{over all } p(u)p(v|u)p(v_1,v_2|v)p(x|v_1,v_2)
	\end{split}
	\right.
	\right\}
	\end{equation}
	and $\mathcal{R}^{\text{old'}}=\mathcal{R}^{\text{old}}$.
\end{lemma}

\begin{IEEEproof}[Proof of Lemma~\ref{lemma 1}]
	The achievability conditions of $\mathcal{R}^{\text{old'}}$ are obtained based on the existing coding scheme for the two-receiver broadcast channel with complementary receiver side information \cite{cref2}. However, the achievability conditions of $\mathcal{R}^{\text{old'}}$ differ slightly from the achievability conditions of $\mathcal{R}^{\text{old}}$  in Theorem~\ref{theorem2}, where constraints (\ref{eq3}.a) and (\ref{eq3}.b) have replaced constraint (\ref{eq1}.a). This is a result of a difference in individual secrecy analysis for the $V^n$ codewords. In Theorem~\ref{theorem2}, both the message segments $M_{1c}$ and $M_{2c}$ are secured by the XOR-ed message segments $M_{1b}\oplus M_{2b}$ and additional randomness $D$. Meanwhile, in Lemma~\ref{lemma 1}, we apply the concept of Carleial-Hellman secrecy coding and treat the respective message segments $M_{1c}$ or $M_{2c}$ as additional randomness in ensuring the individual secrecy of their counterparts. This approach results in the looser constraints (\ref{eq3}.a) and (\ref{eq3}.b). Although $\mathcal{R}^{\text{old}}$ seems to have stricter achievability conditions compared to $\mathcal{R}^{\text{old'}}$, we now show that the two regions are in fact the same.
	
	Applying the Fourier-Motzkin procedure to eliminate the terms $R_a$, $R_b$, $R_{1c}$, $R_{1d}$, $R_{2c}$, $R_{2d}$, $R_D$, $R_{D1}$, $R_{D2}$, $R_{L1}$ and $R_{L2}$ in (\ref{eq3}), we obtain the individual secrecy rate region $\mathcal{R}^{\text{old'}}$, which has the same rate region expression as $\mathcal{R}^{\text{old}}$ in (\ref{eq2}). This gives us the relationship $\mathcal{R}^{\text{old'}}=\mathcal{R}^{\text{old}}$.
\end{IEEEproof}

Using the alternative expression in (\ref{eq3}), we will prove the following theorem.

\begin{theorem}\label{theorem5}
	The simplified coding scheme 1 for the two-receiver discrete memoryless broadcast channel with complementary receiver side information and with a passive eavesdropper achieves an individual secrecy rate region $\mathcal{R}^{\text{new1}}$, where $\mathcal{R}^{\text{new1}}=\mathcal{R}^{\text{old}}$.
\end{theorem}

\begin{IEEEproof}[Proof of Theorem~\ref{theorem5}]
	The proposed simplified coding scheme 1 in Fig.~\ref{fig:6b} can be readily obtained from the existing coding scheme by Chen et al. \cite{cref2} in Fig.~\ref{fig:3b} by setting $M_{ib}=\phi$, for all $i\in\{1,2\}$ in the existing coding scheme. With this change, $R_b=0$, allowing us to rewrite the achievability conditions in (\ref{eq3}) as follows:
	\begin{equation}\label{eq6}
	\left\{
	\begin{split}
	(R_1,R_2)\\ \in \mathbb{R}_+^2
	\end{split}
	\left\vert
	\begin{split}
	&R_1=R_a+R_{1c}+R_{1d},\\
	&R_2=R_a+R_{2c}+R_{2d},\\
	&R_{L1}+R_{L2}>I(V_1;V_2|V),\\
	&R_1+R_D+R_{D1}+R_{L1}<I(U,V,V_1;Y_1),\\
	&R_1-R_a+R_D+R_{D1}+R_{L1}<I(V,V_1;Y_1|U),\\
	&R_{1d}+R_{D1}+R_{L1}<I(V_1;Y_1|V),\\
	&R_2+R_D+R_{D2}+R_{L2}<I(U,V,V_2;Y_2),\\
	&R_2-R_a+R_D+R_{D2}+R_{L2}<I(V,V_2;Y_2|U),\\
	&R_{2d}+R_{D2}+R_{L2}<I(V_2;Y_2|V),\\
	&R_{D1}+R_{L1}\geq I(V_1;Z|V),\\
	&R_{D2}+R_{L2}\geq I(V_2;Z|V),\\
	&R_{1c}+R_D\geq I(V;Z|U),\\
	&R_{2c}+R_D\geq I(V;Z|U),\\
	&R_{L1}+R_{L2}\leq I(V_1;Z|V)+I(V_2;Z|V)\\&-I(V_1,V_2;Z|V),\\
	&R_1\geq 0,R_a\geq 0,R_{1c}\geq 0,R_{1d}\geq 0,\\
	&R_2\geq 0,R_{2c}\geq 0,R_{2d}\geq 0,R_D\geq 0,\\
	&R_{D1}\geq 0,R_{D2}\geq 0,R_{L1}\geq 0,R_{L2}\geq 0\\
	&\text{over all } p(u)p(v|u)p(v_1,v_2|v)p(x|v_1,v_2)
	\end{split}
	\right.
	\right\}
	\end{equation}
	Applying the Fourier-Motzkin procedure to eliminate the terms $R_a$, $R_{1c}$, $R_{1d}$, $R_{2c}$, $R_{2d}$, $R_D$, $R_{D1}$, $R_{D2}$, $R_{L1}$ and $R_{L2}$ in (\ref{eq6}), we obtain $\mathcal{R}^{\text{old}}$. 
\end{IEEEproof}

\subsection{Simplification Step 2}

\begin{figure}[t]
	\centering
	\includegraphics[width=0.35\textwidth]{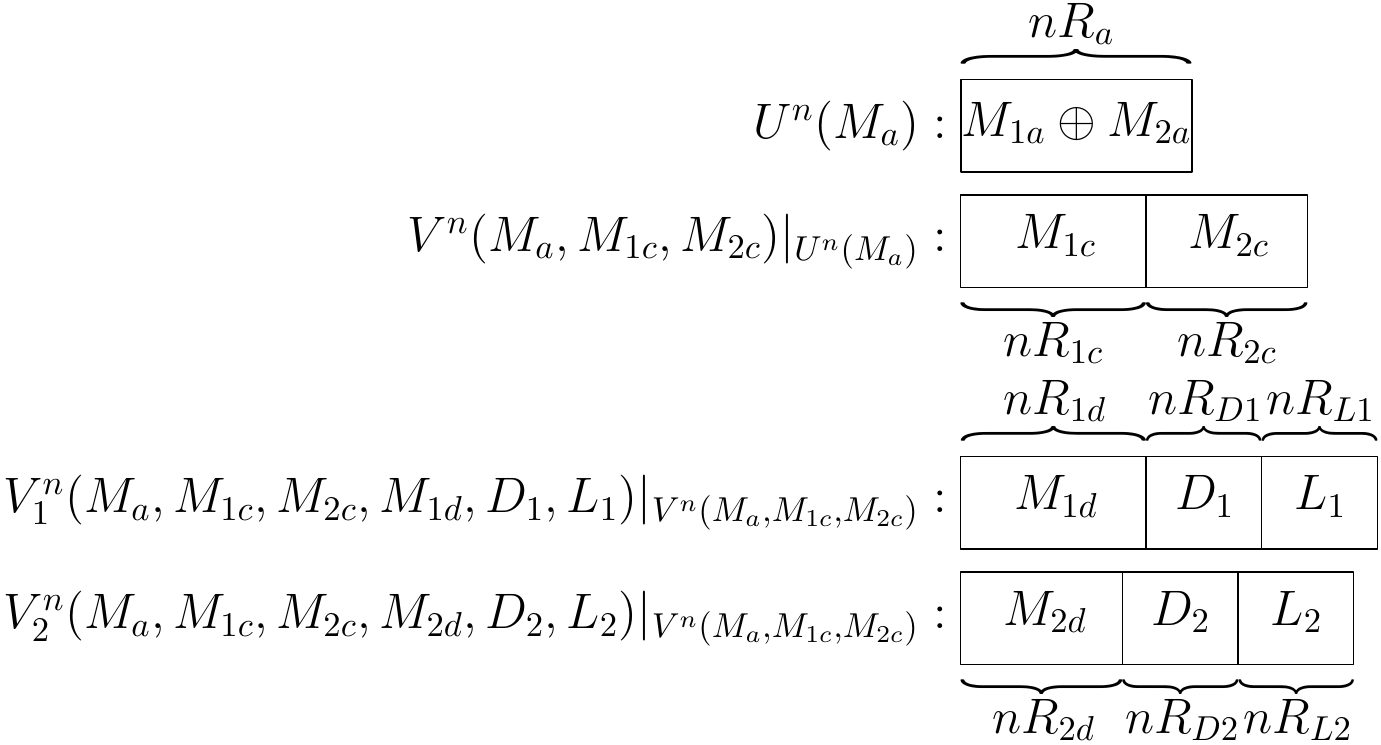}
	\caption{Simplified coding scheme 2.}
	\label{fig:7}
\end{figure}

Referring to the common satellite codeword  $V^n$ of simplified coding scheme 1 in Fig.~\ref{fig:6b}, we recall that Carleial-Hellman secrecy coding \cite{cref3} allows each message to act as a random component which ensures individual secrecy of each another. Intuitively, additional random component should not be necessary in securing messages since each message has provided sufficient randomness. With this concept in mind, we proceed with the removal of the randomness $D$ from the $V^n$ codeword of the previous simplified coding scheme 1. The simplified coding scheme 2 is illustrated in Fig.~\ref{fig:7}. The achievable individual secrecy rate region of this simplified coding scheme 2 is shown in Theorem~\ref{theorem6} and rate region comparison is done through the proofs for Theorem~\ref{theorem7} and Theorem~\ref{theorem8} below.

\begin{theorem}\label{theorem6}
	Define
	\begin{equation}\label{eq7}
	\mathcal{R}_{0}^{\text{new2}}\triangleq
	\left\{
	\begin{split}
	(R_1,R_2)\\ \in \mathbb{R}_+^2
	\end{split}
	\left\vert
	\begin{split}
	&R_1\overset{\text{(a)}}{>}I(V;Z|U),\\
	&R_1\overset{\text{(b)}}{<}I(U;Y_1)+I(V,V_1;Y_1|U)\\&-I(V,V_1;Z|U)+I(V;Z|U),\\
	&R_1-R_2\overset{\text{(c)}}{<}I(V,V_1;Y_1|U)\\&-I(V,V_1;Z|U),\\
	&R_2\overset{\text{(d)}}{>}I(V;Z|U),\\
	&R_2\overset{\text{(e)}}{<}I(U;Y_2)+I(V,V_2;Y_2|U)\\&-I(V,V_2;Z|U)+I(V;Z|U),\\
	&R_2-R_1\overset{\text{(f)}}{<}I(V,V_2;Y_2|U)\\&-I(V,V_2;Z|U)\\
	&\text{over all }p(u)p(v|u)p(v_1,v_2|v)\\&p(x|v_1,v_2)
	\text{ subject to }I(V_1;V_2|V)\overset{\text{(g)}}{<}\\&I(V_1;Z|V)+I(V_2;Z|V)\\&-I(V_1,V_2;Z|V),
	I(V,V_i;Y_i|U)\overset{\text{(h)}}{>}\\&I(V,V_i;Z|U)\text{ and }
	I(V_i;Y_i|V)\overset{\text{(i)}}{>}\\&I(V_i;Z|V)\text{, for all }i\in\{1,2\}.  
	\end{split}
	\right.
	\right\}
	\end{equation}
	and
	\begin{equation}\label{eq8}
	\begin{split}
	\mathcal{R}_{i}^{\text{new2}}\triangleq
	\mathcal{R}_{0}^{\text{new2}} &\text{ without constraints (\ref{eq7}.a), (\ref{eq7}.d) and (\ref{eq7}.g)--(\ref{eq7}.i)}\\ 
	&\text{ but with an additional constraint }V_i=V=U\text{,}\\
	&\text{ for }i\in\{1,2\}.	
	\end{split}
	\end{equation}
	The simplified coding scheme 2 for the two-receiver discrete memoryless broadcast channel with complementary receiver side information and with a passive eavesdropper achieves an individual secrecy rate region $\mathcal{R}^{\text{new2}}$, where
	\vspace{-2pt}
	\begin{equation}\label{eq9}
	\mathcal{R}^{\text{new2}} =
	\mathcal{R}_{0}^{\text{new2}}\cup
	\mathcal{R}_{1}^{\text{new2}}\cup
	\mathcal{R}_{2}^{\text{new2}}.
	\end{equation}
\end{theorem}
\textit{Remark:} By comparing $\mathcal{R}^{\text{old}}$ in (\ref{eq2}) and $\mathcal{R}_{0}^{\text{new2}}$ in (\ref{eq7}), we notice that the latter region is bounded away from the $R_1$ and $R_2$ axes. The origin $(0,0)$ in particular is not in (\ref{eq7}) but is in (\ref{eq2}). This results from the usage of Carleial-Hellman secrecy coding in the $V^n$ common satellite codeword for our proposed simplified coding scheme 2. The principle behind Carleial-Hellman secrecy coding is to let neighbouring message segments act as random components which ensure the individual secrecy of each another. In order to do so, each message segment should reach a certain threshold in randomness to keep the eavesdropper totally ignorant of each message segment. In our case, each of the $M_{1c}$ and $M_{2c}$ message segments has a minimum threshold of $nI(V;Z|U)$ bits to fulfill the role as a random component. This results in constraints (\ref{eq7}.a) and (\ref{eq7}.d), which bound $\mathcal{R}_{0}^{\text{new2}}$ away from the $R_1$ and $R_2$ axes. However, the points along the $R_1$ and $R_2$ axes can be retrieved from $\mathcal{R}_{1}^{\text{new2}}$ and $\mathcal{R}_{2}^{\text{new2}}$. Both $\mathcal{R}_{1}^{\text{new2}}$ and $\mathcal{R}_{2}^{\text{new2}}$ are achievable from a reduced version of the proposed simplified coding scheme 2 as shown in the following proof. 

\begin{IEEEproof}[Proof of Theorem~\ref{theorem6}]
	The proposed simplified coding scheme 2 is a special case of the proposed simplified coding scheme 1 by setting $D=\phi$. Setting $D=\phi$ forces the rate $R_D=0$, hence, we rewrite the achievable individual secrecy rate region in (\ref{eq6}) as
	\begin{equation}\label{eq10}
	\left\{
	\begin{split}
	(R_1,R_2)\\ \in \mathbb{R}_+^2
	\end{split}
	\left\vert
	\begin{split}
	&R_1=R_a+R_{1c}+R_{1d},\\
	&R_2=R_a+R_{2c}+R_{2d},\\
	&R_{L1}+R_{L2}\overset{\text{(a)}}{>}I(V_1;V_2|V),\\
	&R_1+R_{D1}+R_{L1}<I(U,V,V_1;Y_1),\\
	&R_1-R_a+R_{D1}+R_{L1}\overset{\text{(b)}}{<}I(V,V_1;Y_1|U),\\
	&R_{1d}+R_{D1}+R_{L1}\overset{\text{(c)}}{<}I(V_1;Y_1|V),\\
	&R_2+R_{D2}+R_{L2}<I(U,V,V_2;Y_2),\\
	&R_2-R_a+R_{D2}+R_{L2}\overset{\text{(d)}}{<}I(V,V_2;Y_2|U),\\
	&R_{2d}+R_{D2}+R_{L2}\overset{\text{(e)}}{<}I(V_2;Y_2|V),\\
	&R_{D1}+R_{L1}\overset{\text{(f)}}{\geq}I(V_1;Z|V),\\
	&R_{D2}+R_{L2}\overset{\text{(g)}}{\geq}I(V_2;Z|V),\\
	&R_{1c}\overset{\text{(h)}}{\geq}I(V;Z|U),\\
	&R_{2c}\overset{\text{(i)}}{\geq}I(V;Z|U),\\
	&R_{L1}+R_{L2}\overset{\text{(j)}}{\leq}I(V_1;Z|V)+I(V_2;Z|V)\\&-I(V_1,V_2;Z|V),\\
	&R_1\geq 0,R_a\geq 0,R_{1c}\geq 0,R_{1d}\geq 0,\\
	&R_2\geq 0,R_{2c}\geq 0,R_{2d}\geq 0,R_{D1}\geq 0,\\
	&R_{D2}\geq 0,R_{L1}\geq 0,R_{L2}\geq 0\\
	&\text{over all } p(u)p(v|u)p(v_1,v_2|v)p(x|v_1,v_2)
	\end{split}
	\right.
	\right\}
	\end{equation}
	Applying the Fourier-Motzkin procedure to eliminate the terms $R_a$, $R_{1c}$, $R_{1d}$, $R_{2c}$, $R_{2d}$, $R_{D1}$, $R_{D2}$, $R_{L1}$ and $R_{L2}$ in (\ref{eq10}), the region $\mathcal{R}_{0}^{\text{new2}}$ is obtained. 
	
	Note that when $V_i=V=U$, $i\in\{1,2\}$, the proposed scheme reduces to a superposition coding scheme in which only the $U^n$ cloud center codeword and one $V_i^n$ private satellite codeword need to be decoded. The following changes to the simplified coding scheme 2 in Fig.~\ref{fig:7} ensue from the reduction:
	\begin{itemize}
		\item No new message segments are encoded to the $V^n$ codeword, i.e., $M_{1c}=\phi$ and $M_{2c}=\phi$, 
		\item No new message segments are encoded to the reduced $V_i^n$ codeword, i.e., $M_{id}=\phi$ and $D_i=\phi$,
		\item Components of Marton coding are removed, i.e., $L_i=\phi$, for all $i\in\{1,2\}$.
	\end{itemize}
	These changes affect (\ref{eq10}) by
	\begin{itemize}
		\item Removing the rate terms $R_{1c}$, $R_{2c}$, $R_{id}$, $R_{Di}$, $R_{L1}$ and $R_{L2}$ (together with their respective non-negativity constraints), 
		\item Removing achievability constraints of the $V^n$ codeword, i.e., both (\ref{eq10}.b) and (\ref{eq10}.d),
		\item Removing achievability constraints of the reduced $V_i^n$ codeword, i.e., (\ref{eq10}.c) if $i=1$ and (\ref{eq10}.e) if $i=2$,
		\item Removing individual secrecy constraints of the $V^n$ codeword, i.e., both (\ref{eq10}.h) and (\ref{eq10}.i),
		\item Removing individual secrecy constraints of the reduced $V_i^n$ codeword, i.e., (\ref{eq10}.f) if $i=1$ and (\ref{eq10}.g) if $i=2$,
		\item Removing constraints due to Marton coding, i.e., both (\ref{eq10}.a) and (\ref{eq10}.j).
	\end{itemize}
	 Substituting $V_i=V=U$ in the mutual information terms of (\ref{eq10}), we get
	 \begin{equation}\label{eq12}
	 \left\{
	 \begin{split}
	 (R_1,R_2)\\ \in \mathbb{R}_+^2
	 \end{split}
	 \left\vert
	 \begin{split}
	 &R_{\hat{\imath}}=R_a+R_{\hat{\imath}d},\\
	 &R_i=R_a,\\
	 &R_{\hat{\imath}}+R_{D\hat{\imath}}<I(U,V_{\hat{\imath}};Y_{\hat{\imath}}),\\
	 &R_{\hat{\imath}d}+R_{D\hat{\imath}}<I(V_{\hat{\imath}};Y_{\hat{\imath}}|U),\\
	 &R_i<I(U;Y_i),\\
	 &R_{D\hat{\imath}}\geq I(V_{\hat{\imath}};Z|U),\\
	 &R_{\hat{\imath}}\geq 0,R_a\geq 0,R_{\hat{\imath}d}\geq 0, R_i\geq 0,R_{D\hat{\imath}}\geq 0,\\
	 &\text{over all } p(u)p(v_{\hat{\imath}}|u)p(x|v_{\hat{\imath}},u)\\
	 &\text{where }\hat{\imath}\triangleq (i\bmod 2)+1\text{, for }i\in\{1,2\} 
	 \end{split}
	 \right.
	 \right\}
	 \end{equation}
	 By applying the Fourier-Motzkin procedure to eliminate the terms $R_a$, $R_{\hat{\imath}d}$ and $R_{D\hat{\imath}}$ in (\ref{eq12}), we get (\ref{eq7}) with constraints (\ref{eq7}.a), (\ref{eq7}.d) and (\ref{eq7}.g)--(\ref{eq7}.i) removed, giving us the regions $\mathcal{R}_{i}^{\text{new2}}$, for $i\in\{1,2\}$. Taking the union of $\mathcal{R}_{0}^{\text{new2}}$, $\mathcal{R}_{1}^{\text{new2}}$ and $\mathcal{R}_{2}^{\text{new2}}$ will then give us the individual secrecy rate region $\mathcal{R}^{\text{new2}}$ in Theorem~\ref{theorem6}.
\end{IEEEproof}

\begin{figure*}[t]
	\centerline
	{\captionsetup[subfloat]{format=hang,width=0.24\textwidth,justification=raggedright}
		\subfloat[$\mathcal{R}^{\text{old}}$]{\includegraphics[width=0.2\textwidth]{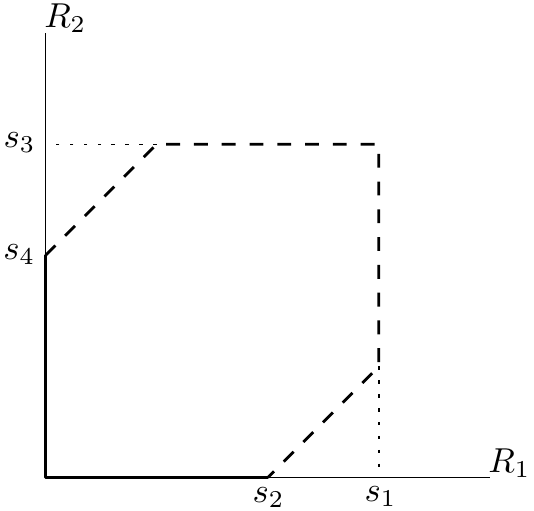}
			\label{fig:8a}}
		\hfil
		\captionsetup[subfloat]{format=hang,width=0.24\textwidth,justification=raggedright}
		\subfloat[$\mathcal{R}_{0}^{\text{new2}}$, \newline $I(V;Z|U)\leq \newline I(V,V_i;Y_i|U)-I(V,V_i;Z|U)$, for all $i\in\{1,2\}$]{\includegraphics[width=0.2\textwidth]{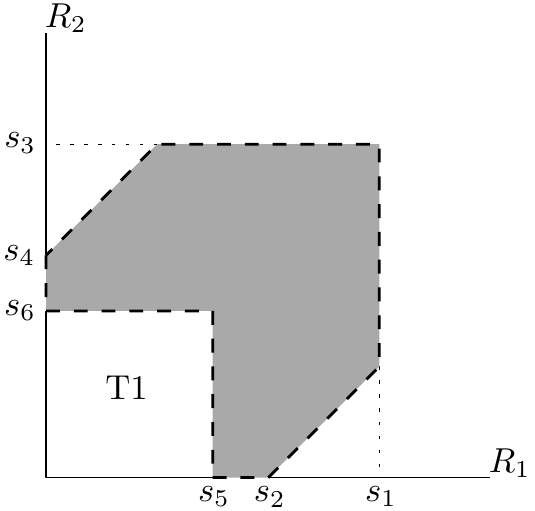}
			\label{fig:8b}}
		\hfil
		\captionsetup[subfloat]{format=hang,width=0.24\textwidth,justification=raggedright}
		\subfloat[$\mathcal{R}_{0}^{\text{new2}}$, \newline $I(V;Z|U)>\newline I(V,V_i;Y_i|U)-I(V,V_i;Z|U)$,\newline  for all $i\in\{1,2\}$]{\includegraphics[width=0.2\textwidth]{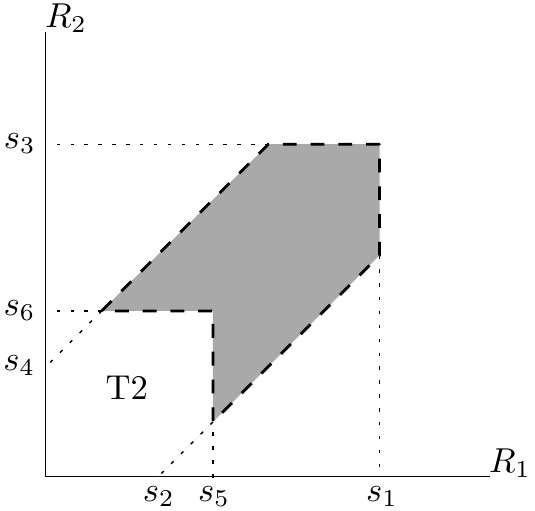}
			\label{fig:8c}}}
	\centerline{\captionsetup[subfloat]{format=hang,width=0.24\textwidth,justification=raggedright}
		\subfloat[$\mathcal{R}_{0}^{\text{new2}}$, \newline $I(V;Z|U)>\newline I(V,V_1;Y_1|U)-I(V,V_1;Z|U)$, \newline $I(V;Z|U)\leq \newline I(V,V_2;Y_2|U)-I(V,V_2;Z|U)$]{\includegraphics[width=0.2\textwidth]{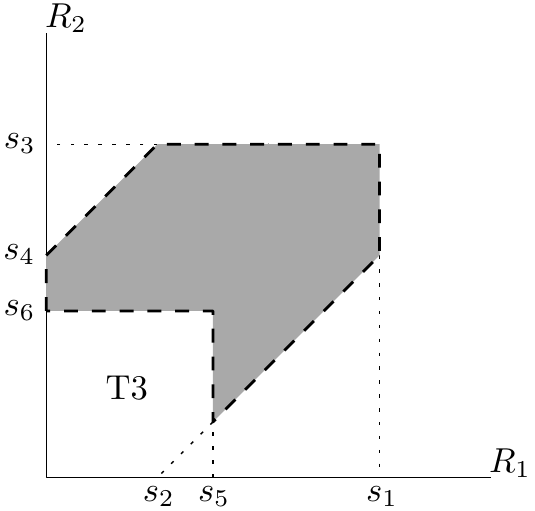}
			\label{fig:8d}}
		\hfil
		\captionsetup[subfloat]{format=hang,width=0.24\textwidth,justification=raggedright}
		\subfloat[$\mathcal{R}_{0}^{\text{new2}}$, \newline $I(V;Z|U)\leq \newline I(V,V_1;Y_1|U)-I(V,V_1;Z|U)$, \newline $I(V;Z|U)>\newline I(V,V_2;Y_2|U)-I(V,V_2;Z|U)$]{\includegraphics[width=0.2\textwidth]{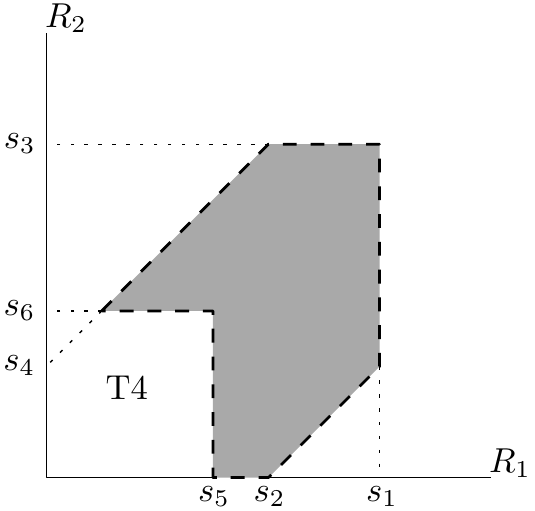}
			\label{fig:8e}}}
	\caption{Plots of individual secrecy rate regions $\mathcal{R}^{\text{new1}}=\mathcal{R}^{\text{old}}$ and $\mathcal{R}_{0}^{\text{new2}}$ (shaded region) for some chosen $p(u)p(v|u)p(v_1,v_2|v)p(x|v_1,v_2)$.}
	\label{fig:8}
\end{figure*}

\begin{theorem}\label{theorem7}
	$\text{co}(\mathcal{R}^{\text{new2}})=\mathcal{R}^{\text{new1}}$.
\end{theorem}

\begin{IEEEproof}[Proof of Theorem~\ref{theorem7}]
	Now, we show that by taking the convex hull of the union of $\mathcal{R}_{0}^{\text{new2}}$, $\mathcal{R}_{1}^{\text{new2}}$ and $\mathcal{R}_{2}^{\text{new2}}$, we will be able to achieve the region $\mathcal{R}^{\text{new1}}$. As a start, we plot the rate region $\mathcal{R}^{\text{new1}}$ as illustrated in Fig.~\ref{fig:8a}. It can be observed that $\mathcal{R}_{0}^{\text{new2}}$ has two additional constraints (\ref{eq7}.a) and (\ref{eq7}.d) compared to $\mathcal{R}^{\text{new1}}$. By comparing RHS of (\ref{eq7}.a) and (\ref{eq7}.d) to RHS of (\ref{eq7}.c) and (\ref{eq7}.f), $\mathcal{R}_{0}^{\text{new2}}$ can be plotted for four cases as shown in Fig.~\ref{fig:8b} to Fig.~\ref{fig:8e}. For all plots in Fig.~\ref{fig:8}, we have labelled
	\begin{flalign*}
	&s_1=(A-\varepsilon,0)\text{, } s_2=(B-\varepsilon,0)\text{, } s_3=(0,C-\varepsilon)\text{, }&\\ &s_4=(0,D-\varepsilon)\text{, } s_5=(E-\varepsilon,0)\text{, }
	s_6=(0,E-\varepsilon)&
	\end{flalign*}
	where
	\begin{flalign*}
	&A\triangleq I(U;Y_1)+I(V,V_1;Y_1|U)-I(V,V_1;Z|U)+I(V;Z|U)\text{,}&\\
	&B\triangleq I(V,V_1;Y_1|U)-I(V,V_1;Z|U)\text{,}&\\ &C\triangleq I(U;Y_2)+I(V,V_2;Y_2|U)-I(V,V_2;Z|U)+I(V;Z|U)\text{,}&\\
	&D\triangleq I(V,V_2;Y_2|U)-I(V,V_2;Z|U)\text{,}&\\ &E\triangleq I(V;Z|U)&
	\end{flalign*}
	for some arbitrarily small $\varepsilon$, $\varepsilon>0$. Note that $E\leq A$ and $E\leq C$ at all times. The proof of $E\leq A$ is as follows:
	\begin{align*}
	A&=I(U;Y_1)+I(V,V_1;Y_1|U)-I(V,V_1;Z|U)+I(V;Z|U)\\
	&\overset{\text{(a)}}{\geq}I(U;Y_1)+I(V;Z|U)\\
	&\geq I(V;Z|U)\\
	&=E
	\end{align*}
	where (a) follows from (\ref{eq2}.b) and (\ref{eq7}.h). The same proof applies to $E\leq C$.
	
	Referring to Fig.~\ref{fig:8b}, when $I(V;Z|U)\leq I(V,V_i;Y_i|U)-I(V,V_i;Z|U)$, for all $i\in\{1,2\}$, the additional constraints (\ref{eq7}.a) and (\ref{eq7}.d) result in the excluded square region T1. Meanwhile, in Fig.~\ref{fig:8c}, when $I(V;Z|U)>I(V,V_i;Y_i|U)-I(V,V_i;Z|U)$, for all $i\in\{1,2\}$, (\ref{eq7}.a) and (\ref{eq7}.d) form a hexagonal cut and result in the excluded region T2. On the other hand, for the remaining cases in Fig.~\ref{fig:8d} and Fig.~\ref{fig:8e}, (\ref{eq7}.a) and (\ref{eq7}.d) form a pentagonal cut and result in the excluded regions T3 and T4 respectively. In order to retrieve the excluded rate points in regions T1 to T4, it is essential to recover point $s_2$ and point $s_4$. To recover point $s_2$, we first consider the random variables $(U,V,V_1)$ that are arbitrarily correlated. Next, from $\mathcal{R}_{2}^{\text{new2}}$ in (\ref{eq8}), we set $V'_2=V'=U'=U$ and $V'_1=(V,V_1)$ such that $(U',V',V'_1,V'_1,V'_2,X,Y_1,Y_2,Z)$ satisfy $p(u')p(v'|u')p(v'_1,v'_2|v')p(x|v'_1,v'_2)p(y_1,y_2,z|x)$ and the constraint $V'_2=V'=U'$. The substitutions are as follows:\\
	From (\ref{eq7}.b),\\
	\begin{flalign*}
	R_1<&I(U';Y_1)+I(V',V'_1;Y_1|U')-I(V',V'_1;Z|U')&\\&+I(V';Z|U')&\\
	=&I(U;Y_1)+I(U,V,V_1;Y_1|U)-I(U,V,V_1;Z|U)&\\&+I(U;Z|U)&\\
	=&I(U;Y_1)+I(V,V_1;Y_1|U)-I(V,V_1;Z|U)&
	\end{flalign*}
	From (\ref{eq7}.c),\\
	\begin{flalign*}
	R_1-R_2<&I(V',V'_1;Y_1|U')-I(V',V'_1;Z|U')&\\
	=&I(U,V,V_1;Y_1|U)-I(U,V,V_1;Z|U)&\\
	=&I(V,V_1;Y_1|U)-I(V,V_1;Z|U)&
	\end{flalign*}
	From (\ref{eq7}.e),\\
	\begin{flalign*}
	R_2<&I(U';Y_2)+I(V',V'_2;Y_2|U')-I(V',V'_2;Z|U')&\\&+I(V';Z|U')&\\
	=&I(U;Y_2)+I(U;Y_2|U)-I(U;Z|U)+I(U;Z|U)&\\
	=&I(U;Y_2)&
	\end{flalign*}
	From (\ref{eq7}.f),\\
	\begin{flalign*}
	R_2-R_1<&I(V',V'_2;Y_2|U')-I(V',V'_2;Z|U')&\\
	=&I(U;Y_2|U)-I(U;Z|U)&\\
	\text{And hence, }\\
	R_2<&R_1&
	\end{flalign*}
	Note that we have been using constraints from (\ref{eq7}) for the substitutions since (\ref{eq8}) is expressed in terms of (\ref{eq7}). This series of substitutions allows us to obtain a subregion $\mathcal{R}_{2}^{\text{new2'}}$:
	\begin{equation}\label{eq11}
	\mathcal{R}_{2}^{\text{new2'}}\triangleq
	\left\{
	\begin{split}
	(R_1,R_2)\\ \in \mathbb{R}_+^2
	\end{split}
	\left\vert
	\begin{split}
	&R_1\overset{\text{(a)}}{<}I(U;Y_1)+I(V,V_1;Y_1|U)\\&-I(V,V_1;Z|U),\\
	&R_1-R_2\overset{\text{(b)}}{<}I(V,V_1;Y_1|U)\\&-I(V,V_1;Z|U),\\
	&R_2\overset{\text{(c)}}{<}I(U;Y_2),\\
	&R_2\overset{\text{(d)}}{<}R_1\\
	&\text{over all }p(u,v,v_1)\\&p(x|u,v,v_1).
	\end{split}
	\right.
	\right\}
	\end{equation}
	From (\ref{eq2}), point $s_2$ has the coordinates $(I(V,V_1;Y_1|U)-I(V,V_1;Z|U)-\varepsilon,0)$ for some arbitrarily small $\varepsilon$, $\varepsilon>0$ where $(U,V,V_1,V_2)$ satisfy $p(u)p(v|u)p(v_1,v_2|v)p(x|v_1,v_2)$ and constraints (\ref{eq2}.a)--(\ref{eq2}.c). From here, we observe that the coordinates of $s_2$ satisfies (\ref{eq11}.a)--(\ref{eq11}.d). This shows that $s_2$ is recoverable and lies in $\mathcal{R}_{2}^{\text{new2}}$ since $\mathcal{R}_{2}^{\text{new2'}} \subseteq \mathcal{R}_{2}^{\text{new2}}$. Likewise, to recover point $s_4$, we again consider the random variables $(U,V,V_2)$ that are arbitrarily correlated. Next, from $\mathcal{R}_{1}^{\text{new2}}$ in (\ref{eq8}), we can set $V'_1=V'=U'=U$ and $V'_2=(V,V_2)$ such that $(U',V',V'_1,V'_1,V'_2,X,Y_1,Y_2,Z)$ satisfy $p(u')p(v'|u')p(v'_1,v'_2|v')p(x|v'_1,v'_2)p(y_1,y_2,z|x)$ and the constraint $V'_1=V'=U'$. These arguments show that $s_2\in \mathcal{R}_{2}^{\text{new2}}$ and $s_4\in \mathcal{R}_{1}^{\text{new2}}$. Noting that the point $(0,0)\in \mathcal{R}_{i}^{\text{new2}}$, for all $i\in\{1,2\}$, hence, by time-sharing, we take co$(\mathcal{R}_{0}^{\text{new2}}\cup \mathcal{R}_{1}^{\text{new2}}\cup \mathcal{R}_{2}^{\text{new2}})$ to achieve $\mathcal{R}^{\text{new1}}$. 
\end{IEEEproof}

\begin{theorem}\label{theorem8}
	$\text{co}(\mathcal{R}^{\text{new2}})=\mathcal{R}^{\text{new2}}$.
\end{theorem}

\begin{IEEEproof}[Proof of Theorem~\ref{theorem8}]
	
	In the proof for Theorem~\ref{theorem7}, we show that through time-sharing, we take $\text{co}(\mathcal{R}^{\text{new2}})$ to obtain $\mathcal{R}^{\text{new1}}$. Time-sharing, however, is redundant, i.e.,  $\mathcal{R}^{\text{new2}}=\mathcal{R}^{\text{new1}}$, if the following conditions are met:
	\begin{enumerate}
		\item $\mathcal{R}_{0}^{\text{new2}}$, $\mathcal{R}_{1}^{\text{new2}}$ and $\mathcal{R}_{2}^{\text{new2}}$ are convex.
		\item $\mathcal{R}_{0}^{\text{new2}}$ contains all the interior points of $\mathcal{R}^{\text{new1}}$.
		\item $\mathcal{R}_{1}^{\text{new2}}$ and $\mathcal{R}_{2}^{\text{new2}}$ contain all boundary points of $\mathcal{R}^{\text{new1}}$ that are excluded by $\mathcal{R}_{0}^{\text{new2}}$.
	\end{enumerate}
	Fulfilling all three conditions above allows us to recover all points of $\mathcal{R}^{\text{new1}}$ by taking the union of $\mathcal{R}_{0}^{\text{new2}}$, $\mathcal{R}_{1}^{\text{new2}}$ and $\mathcal{R}_{2}^{\text{new2}}$, i.e., $\mathcal{R}^{\text{new2}}$.
	
	In order to proof condition 1, we first note that the regions $\mathcal{R}_{0}^{\text{new2}}$, $\mathcal{R}_{1}^{\text{new2}}$ and $\mathcal{R}_{2}^{\text{new2}}$ can be convexified by introducing a time-sharing variable $Q$ which is independent of $(U,V,V_1,V_2,X,Y_1,Y_2,Z)$. Alternatively, by defining $U'=(U,Q)$, $V'=(V,Q)$, $V'_1=(V_1,Q)$ and $V'_2=(V_2,Q)$, we naturally include the time-sharing variable, hence, showing that $\mathcal{R}_{0}^{\text{new2}}$, $\mathcal{R}_{1}^{\text{new2}}$ and $\mathcal{R}_{2}^{\text{new2}}$ are convex.
	
	Recalling our earlier discussion, we understand that $\mathcal{R}_{0}^{\text{new2}}$ is bounded away from the $R_1$ and $R_2$ axes due to constraints (\ref{eq7}.a) and (\ref{eq7}.d). As a result, in order to fulfill condition 2, we need to show that $\mathcal{R}_{0}^{\text{new2}}$ can go arbitrarily close to the $R_1$ and $R_2$ axes, i.e., $\mathcal{R}_{0}^{\text{new2}}$ approaches the points $(0,0)$, $s_2$ and $s_4$ in Fig.~\ref{fig:8a}. 
	
	Now, we show that $\mathcal{R}_{0}^{\text{new2}}$ goes arbitrarily close to the $R_1$ axis by proving that the points $(\varepsilon,\varepsilon)$ and $(B-\varepsilon,\varepsilon)$ are in $\mathcal{R}_{0}^{\text{new2}}$ for any arbitrarily small $\varepsilon$, $\varepsilon>0$. Using an argument similar to time-sharing, let $\mathcal{R}_{0}^{\text{new2}}$ be achievable from two sets of random variables $(U_\alpha,V_\alpha,V_{1\alpha},V_{2\alpha})$ and $(U_\beta,V_\beta,V_{1\beta},V_{2\beta})$ through an auxiliary random variable $Q$ 
	with probability mass function $p_{}(q)$ given by:
	\begin{flalign*}
	p_{}(q)=&
	\biggl\{
	\begin{aligned}
	\gamma &\text{\qquad if }q=\alpha\\ 
	1-\gamma &\text{\qquad if }q=\beta
	\end{aligned}&
	\end{flalign*}
	where $\alpha$ and $\beta$ are integer values taken by $Q$ at different instances of time-sharing and $0< \gamma \leq 1$. 
	
	Considering a new set of random variables $(U^*,V^*,V^*_1,V^*_2,X^*,Y^*_1,Y^*_2,Z^*)$ which satisfies some  $p(u^*)p(v^*|u^*)p(v^*_1,v^*_2|v^*)p(x^*|v^*_1,v^*_2)p(y^*_1,y^*_2,z^*|x^*)$, we define:
	\begin{itemize}[leftmargin=*]
		\item $U_\alpha=U^*,V_\alpha=V^*\text{ and }V_{i\alpha}=V^*_i,\text{for all }i\in\{1,2\}$ such that $(U_\alpha,V_\alpha,V_{1\alpha},V_{2\alpha})$ satisfy $p(u_\alpha)p(v_\alpha|u_\alpha)p(v_{1\alpha},v_{2\alpha}|v_\alpha)p(x^*|v_{1\alpha},v_{2\alpha})p(y^*_1,y^*_2,z^*|x^*)$,
		\item $V_{2\beta}=V_\beta=U_\beta=U^*\text{ and }V_{1\beta}=(V^*,V^*_1)$ such that $(U_\beta,V_\beta,V_{1\beta},V_{2\beta})$ satisfy $p(u_\beta)p(v_\beta|u_\beta)p(v_{1\beta},v_{2\beta}|v_\beta)p(x^*|v_{1\beta},v_{2\beta})p(y^*_1,y^*_2,z^*|x^*)$ and the constraint $V_{2\beta}=V_\beta=U_\beta$.
	\end{itemize}
	
	In order to incorporate the auxiliary random variable $Q$ into (\ref{eq7}), we next define additional random variables $U'=(U,Q)$, $V'=(V,Q)$, $V'_1=(V_1,Q)$ and $V'_2=(V_2,Q)$ such that $(U',V',V'_1,V'_2,X,Y_1,Y_2,Z)$ has the pmf $p(u')p(v'|u')p(v'_1,v'_2|v')p(x|v'_1,v'_2)p(y_1,y_2,z|x)$ assigned as follows:
	\begin{flalign*}
	p_{}(u')=&p_{}(u,q)&\\
	=&p_{}(q)p_{}(u|q)&\\
	=&\biggl\{
	\begin{aligned}
	p_{}(\alpha)p_{}(u|\alpha) &\text{\qquad if }q=\alpha\\ 
	p_{}(\beta)p_{}(u|\beta) &\text{\qquad if }q=\beta
	\end{aligned}&\\
	\overset{\text{assign}}{=}&\biggl\{
	\begin{aligned}
	\gamma p(u_\alpha) &\text{\qquad if }q=\alpha\\ 
	(1-\gamma) p(u_\beta) &\text{\qquad if }q=\beta
	\end{aligned}&
	\end{flalign*}
	\begin{flalign*}
	p_{}(v'|u')=&p_{}(v,q|u,q)&\\
	=&p_{}(q)p_{}(v|u,q)&\\
	=&\biggl\{
	\begin{aligned}
	p_{}(\alpha)p_{}(v|u,\alpha) &\text{\qquad if }q=\alpha\\ 
	p_{}(\beta)p_{}(v|u,\beta) &\text{\qquad if }q=\beta
	\end{aligned}&\\
	\overset{\text{assign}}{=}&\biggl\{
	\begin{aligned}
	\gamma p(v_\alpha|u_\alpha) &\text{\qquad if }q=\alpha\\ 
	(1-\gamma) p(v_\beta|u_\beta) &\text{\qquad if }q=\beta
	\end{aligned}&
	\end{flalign*}
	\begin{flalign*}
	p_{}(v'_1|v')=&p_{}(v_1,q|v,q)&\\
	=&p_{}(q)p_{}(v_1|v,q)&\\
	=&\biggl\{
	\begin{aligned}
	p_{}(\alpha)p_{}(v_1|v,\alpha) &\text{\qquad if }q=\alpha\\ 
	p_{}(\beta)p_{}(v_1|v,\beta) &\text{\qquad if }q=\beta
	\end{aligned}&\\
	\overset{\text{assign}}{=}&\biggl\{
	\begin{aligned}
	\gamma p(v_{1\alpha}|v_\alpha) &\text{\qquad if }q=\alpha\\ 
	(1-\gamma) p(v_{1_\beta}|v_\beta) &\text{\qquad if }q=\beta
	\end{aligned}&
	\end{flalign*}
	\begin{flalign*}
	p_{}(v'_2|v')=&p_{}(v_2,q|v,q)&\\
	=&p_{}(q)p_{}(v_2|v,q)&\\
	=&\biggl\{
	\begin{aligned}
	p_{}(\alpha)p_{}(v_2|v,\alpha) &\text{\qquad if }q=\alpha\\ 
	p_{}(\beta)p_{}(v_2|v,\beta) &\text{\qquad if }q=\beta
	\end{aligned}&\\
	\overset{\text{assign}}{=}&\biggl\{
	\begin{aligned}
	\gamma p(v_{2\alpha}|v_\alpha) &\text{\qquad if }q=\alpha\\ 
	(1-\gamma) p(v_{2\beta}|v_\beta) &\text{\qquad if }q=\beta
	\end{aligned}&
	\end{flalign*}
	\begin{flalign*}
	p_{}(x|v'_1,v'_2)=&p_{}(x|v_1,v_2,q)&\\
	=&\biggl\{
	\begin{aligned}
	p_{}(x|v_1,v_2,\alpha) &\text{\qquad if }q=\alpha\\ 
	p_{}(x|v_1,v_2,\beta) &\text{\qquad if }q=\beta
	\end{aligned}&\\
	\overset{\text{assign}}{=}&\biggl\{
	\begin{aligned}
	p(x^*|v_{1\alpha},v_{2\alpha}) &\text{\qquad if }q=\alpha\\ 
	p(x^*|v_{1\beta},v_{2\beta}) &\text{\qquad if }q=\beta
	\end{aligned}&
	\end{flalign*}
	
	Substituting these properties into (\ref{eq7}), we have:
	\\From (\ref{eq7}.a) and (\ref{eq7}.d),
	\begin{flalign}\label{eq13}
	R_1>&I(V';Z|U')&\nonumber\\
	=&I(V,Q;Z|U,Q)&\nonumber\\
	=&I(V;Z|U,Q)&\nonumber\\
	=&\sum_q p(q)I(V;Z|U,Q=q)&\nonumber\\
	=&p(\alpha)I(V_\alpha;Z^*|U_\alpha)+p(\beta)I(V_\beta;Z^*|U_\beta)&\nonumber\\
	=&\gamma I(V^*;Z^*|U^*)+(1-\gamma)I(U^*;Z^*|U^*)&\nonumber\\
	=&\gamma I(V^*;Z^*|U^*)
	\end{flalign}
	\begin{flalign}\label{eq14}
	R_2>&I(V';Z|U')&\nonumber\\
	=&\gamma I(V^*;Z^*|U^*)
	\end{flalign}
	From (\ref{eq7}.b),\\
	\begin{flalign}\label{eq15}
	R_1<&I(U';Y_1)+I(V',V'_1;Y_1|U')-I(V',V'_1;Z|U')&\nonumber\\&+I(V';Z|U')&\nonumber\\
	=&I(U,Q;Y_1)+I(V,V_1,Q;Y_1|U,Q)&\nonumber\\&-I(V,V_1,Q;Z|U,Q)+I(V,Q;Z|U,Q)&\nonumber\\
	=&I(Q;Y_1)+I(U;Y_1|Q)+I(V,V_1;Y_1|U,Q)&\nonumber\\&-I(V,V_1;Z|U,Q)+I(V;Z|U,Q)&\nonumber\\
	=&\sum_q p(q) [I(U;Y_1|Q=q)+I(V,V_{1};Y_1|U,Q=q)&\nonumber\\&-I(V,V_{1};Z|U,Q=q+I(V;Z|U,Q=q)]+I(Q;Y_1)&\nonumber\\
	=&p(\alpha) [I(U_\alpha;Y^*_1)+I(V_\alpha,V_{1\alpha};Y^*_1|U_\alpha)-I(V_\alpha,V_{1\alpha};Z^*|U_\alpha)&\nonumber\\&+I(V_\alpha;Z^*|U_\alpha)]+p(\beta) [I(U_\beta;Y^*_1)+I(V_\beta,V_{1\beta};Y^*_1|U_\beta)&\nonumber\\&-I(V_\beta,V_{1\beta};Z^*|U_\beta)+I(V_\beta;Z^*|U_\beta)]+I(Q;Y_1)&\nonumber\\
	=&\gamma [I(U^*;Y^*_1)+I(V^*,V^*_{1};Y^*_1|U^*)-I(V^*,V^*_{1};Z^*|U^*)&\nonumber\\&+I(V^*;Z^*|U^*)]+(1-\gamma) [I(U^*;Y^*_1)&\nonumber\\&+I(U^*,V^*,V^*_{1};Y^*_1|U^*)-I(U^*,V^*,V^*_{1};Z^*|U^*)&\nonumber\\&+I(U^*;Z^*|U^*)]+I(Q;Y_1)&\nonumber\\
	=&\gamma [I(U^*;Y^*_1)+I(V^*,V^*_{1};Y^*_1|U^*)-I(V^*,V^*_{1};Z^*|U^*)&\nonumber\\&+I(V^*;Z^*|U^*)]+(1-\gamma) [I(U^*;Y^*_1)&\nonumber\\&+I(V^*,V^*_{1};Y^*_1|U^*)-I(V^*,V^*_{1};Z^*|U^*)]+I(Q;Y_1)&\nonumber\\
	=&I(U^*;Y^*_1)+I(V^*,V^*_{1};Y^*_1|U^*)-I(V^*,V^*_{1};Z^*|U^*)&\nonumber\\&+\gamma I(V^*;Z^*|U^*)+I(Q;Y_1)&
	\end{flalign}
	From (\ref{eq7}.c),\\
	\begin{flalign}\label{eq16}
	R_1-R_2<&I(V',V'_1;Y_1|U')-I(V',V'_1;Z|U')&\nonumber\\
	=&I(V,V_1,Q;Y_1|U,Q)-I(V,V_1,Q;Z|U,Q)&\nonumber\\
	=&I(V,V_1;Y_1|U,Q)-I(V,V_1;Z|U,Q)&\nonumber\\
	=&\sum_q p(q) [I(V,V_{1};Y_1|U,Q=q)&\nonumber\\&-I(V,V_{1};Z|U,Q=q)]&\nonumber\\
	=&p(\alpha) [I(V_\alpha,V_{1\alpha};Y^*_1|U_\alpha)-I(V_\alpha,V_{1\alpha};Z^*|U_\alpha)]&\nonumber\\&+p(\beta) [I(V_\beta,V_{1\beta};Y^*_1|U_\beta)-I(V_\beta,V_{1\beta};Z^*|U_\beta)]&\nonumber\\
	=&\gamma [I(V^*,V^*_{1};Y^*_1|U^*)-I(V^*,V^*_{1};Z^*|U^*)]&\nonumber\\&+(1-\gamma) [I(U^*,V^*,V^*_{1};Y^*_1|U^*)&\nonumber\\&-I(U^*,V^*,V^*_{1};Z^*|U^*)]&\nonumber\\
	=&\gamma [I(V^*,V^*_{1};Y^*_1|U^*)-I(V^*,V^*_{1};Z^*|U^*)]&\nonumber\\&+(1-\gamma) [I(V^*,V^*_{1};Y^*_1|U^*)-I(V^*,V^*_{1};Z^*|U^*)]&\nonumber\\
	=&I(V^*,V^*_{1};Y^*_1|U^*)-I(V^*,V^*_{1};Z^*|U^*)]&
	\end{flalign}
	From (\ref{eq7}.e),\\
	\begin{flalign}\label{eq17}
	R_2<&I(U';Y_2)+I(V',V'_2;Y_2|U')-I(V',V'_2;Z|U')&\nonumber\\&+I(V';Z|U')&\nonumber\\
	=&I(U,Q;Y_2)+I(V,V_2,Q;Y_2|U,Q)&\nonumber\\&-I(V,V_2,Q;Z|U,Q)+I(V,Q;Z|U,Q)&\nonumber\\
	=&I(Q;Y_2)+I(U;Y_2|Q)+I(V,V_2;Y_2|U,Q)&\nonumber\\&-I(V,V_2;Z|U,Q)+I(V;Z|U,Q)&\nonumber\\
	=&\sum_q p(q) [I(U;Y_2|Q=q)+I(V,V_{2};Y_2|U,Q=q)&\nonumber\\&-I(V,V_{2};Z|U,Q=q)+I(V;Z|U,Q=q)]+I(Q;Y_2)&\nonumber\\
	=&p(\alpha) [I(U_\alpha;Y^*_2)+I(V_\alpha,V_{2\alpha};Y^*_2|U_\alpha)&\nonumber\\&-I(V_\alpha,V_{2\alpha};Z^*|U_\alpha)+I(V_\alpha;Z^*|U_\alpha)]&\nonumber\\&+p(\beta) [I(U_\beta;Y^*_2)+I(V_\beta,V_{2\beta};Y^*_2|U_\beta)&\nonumber\\&-I(V_\beta,V_{2\beta};Z^*|U_\beta)+I(V_\beta;Z^*|U_\beta)]+I(Q;Y_2)&\nonumber\\
	=&\gamma [I(U^*;Y^*_2)+I(V^*,V^*_{2};Y^*_2|U^*)-I(V^*,V^*_{2};Z^*|U^*)&\nonumber\\&+I(V^*;Z^*|U^*)]+(1-\gamma) [I(U^*;Y^*_2)+I(U^*;Y^*_2|U^*)&\nonumber\\&-I(U^*;Z^*|U^*)+I(U^*;Z^*|U^*)]+I(Q;Y_2)&\nonumber\\
	=&\gamma [I(U^*;Y^*_2)+I(V^*,V^*_{2};Y^*_2|U^*)-I(V^*,V^*_{2};Z^*|U^*)&\nonumber\\&+I(V^*;Z^*|U^*)]+(1-\gamma) I(U^*;Y^*_2)+I(Q;Y_2)&\nonumber\\
	=&I(U^*;Y^*_2)+\gamma [I(V^*,V^*_{2};Y^*_2|U^*)-I(V^*,V^*_{2};Z^*|U^*)&\nonumber\\&+I(V^*;Z^*|U^*)]+I(Q;Y_2)&
	\end{flalign}
	From (\ref{eq7}.f),\\
	\begin{flalign}\label{eq18}
	R_2-R_1<&I(V',V'_2;Y_2|U')-I(V',V'_2;Z|U')&\nonumber\\
	=&I(V,V_2,Q;Y_2|U,Q)-I(V,V_2,Q;Z|U,Q)&\nonumber\\
	=&I(V,V_2;Y_2|U,Q)-I(V,V_2;Z|U,Q)&\nonumber\\
	=&\sum_q p(q) [I(V,V_{2};Y_2|U,Q=q)&\nonumber\\&-I(V,V_{2};Z|U,Q=q)]&\nonumber\\
	=&p(\alpha) [I(V_\alpha,V_{2\alpha};Y^*_2|U_\alpha)-I(V_\alpha,V_{2\alpha};Z^*|U_\alpha)]&\nonumber\\&+p(\beta) [I(V_\beta,V_{2\beta};Y^*_2|U_\beta)-I(V_\beta,V_{2\beta};Z^*|U_\beta)]&\nonumber\\
	=&\gamma [I(V^*,V^*_{2};Y^*_2|U^*)-I(V^*,V^*_{2};Z^*|U^*)]&\nonumber\\&+(1-\gamma)[I(U^*;Y^*_2|U^*)-I(U^*;Z^*|U^*)]&\nonumber\\
	=&\gamma [I(V^*,V^*_{2};Y^*_2|U^*)-I(V^*,V^*_{2};Z^*|U^*)]&
	\end{flalign}
	From (\ref{eq7}.g),\\
	\begin{flalign}\label{eq19}
	0<&I(V'_1;Z|V')+I(V'_2;Z|V')-I(V'_1,V'_2;Z|V')&\nonumber\\&-I(V'_1;V'_2|V')&\nonumber\\
	=&I(V_1,Q;Z|V,Q)+I(V_2,Q;Z|V,Q)&\nonumber\\&-I(V_1,V_2,Q;Z|V,Q)-I(V_1,Q;V_2,Q|V,Q)&\nonumber\\
	=&I(V_1;Z|V,Q)+I(V_2;Z|V,Q)-I(V_1,V_2;Z|V,Q)&\nonumber\\&-I(V_1;V_2|V,Q)&\nonumber\\
	=&\sum_q p(q) [I(V_{1};Z|V,Q=q)+I(V_{2};Z|V,Q=q)&\nonumber\\&-I(V_{1},V_{2};Z|V,Q=q)-I(V_{1};V_{2}|V,Q=q)]&\nonumber\\
	=&p(\alpha) [I(V_{1\alpha};Z^*|V_\alpha)+I(V_{2\alpha};Z^*|V_\alpha)-I(V_{1\alpha},V_{2\alpha};Z^*|V_\alpha)&\nonumber\\&-I(V_{1\alpha};V_{2\alpha}|V_\alpha)]+p(\beta) [I(V_{1\beta};Z^*|V_\beta)+I(V_{2\beta};Z^*|V_\beta)&\nonumber\\&-I(V_{1\beta},V_{2\beta};Z^*|V_\beta)-I(V_{1\beta};V_{2\beta}|V_\beta)]&\nonumber\\
	=&\gamma [I(V^*_{1};Z^*|V^*)+I(V^*_{2};Z^*|V^*)-I(V^*_{1},V^*_{2};Z^*|V^*)&\nonumber\\&-I(V^*_{1};V^*_{2}|V^*)]+(1-\gamma) [I(V^*,V^*_{1};Z^*|U^*)&\nonumber\\&+I(U^*;Z^*|U^*)-I(V^*,V^*_{1},U^*;Z^*|U^*)&\nonumber\\&-I(V^*,V^*_{1};U^*|U^*)]&\nonumber\\
	=&\gamma [I(V^*_{1};Z^*|V^*)+I(V^*_{2};Z^*|V^*)-I(V^*_{1},V^*_{2};Z^*|V^*)&\nonumber\\&-I(V^*_{1};V^*_{2}|V^*)]&
	\end{flalign}
	From (\ref{eq7}.h),\\
	\begin{flalign}\label{eq20}
	0<&I(V',V'_1;Y_1|U')-I(V',V'_1;Z|U')&\nonumber\\
	=&I(V,V_1,Q;Y_1|U,Q)-I(V,V_1,Q;Z|U,Q)&\nonumber\\
	=&I(V,V_1;Y_1|U,Q)-I(V,V_1;Z|U,Q)&\nonumber\\
	=&\sum_q p(q) [I(V,V_{1};Y_1|U,Q=q)-I(V,V_{1};Z|U,Q=q)]&\nonumber\\
	=&p(\alpha) [I(V_\alpha,V_{1\alpha};Y^*_1|U_\alpha)-I(V_\alpha,V_{1\alpha};Z^*|U_\alpha)]&\nonumber\\&+p(\beta) [I(V_\beta,V_{1\beta};Y^*_1|U_\beta)-I(V_\beta,V_{1\beta};Z^*|U_\beta)]&\nonumber\\
	=&\gamma [I(V^*,V^*_{1};Y^*_1|U^*)-I(V^*,V^*_{1};Z^*|U^*)]&\nonumber\\&+(1-\gamma) [I(U^*,V^*,V^*_{1};Y^*_1|U^*)-I(U^*,V^*,V^*_{1};Z^*|U^*)]&\nonumber\\
	=&\gamma [I(V^*,V^*_{1};Y^*_1|U^*)-I(V^*,V^*_{1};Z^*|U^*)]&\nonumber\\&+(1-\gamma) [I(V^*,V^*_{1};Y^*_1|U^*)-I(V^*,V^*_{1};Z^*|U^*)]&\nonumber\\
	=&I(V^*,V^*_{1};Y^*_1|U^*)-I(V^*,V^*_{1};Z^*|U^*)]&
	\end{flalign}
	\begin{flalign}\label{eq21}
	0<&I(V',V'_2;Y_2|U')-I(V',V'_2;Z|U')&\nonumber\\
	=&I(V,V_2,Q;Y_2|U,Q)-I(V,V_2,Q;Z|U,Q)&\nonumber\\
	=&I(V,V_2;Y_2|U,Q)-I(V,V_2;Z|U,Q)&\nonumber\\
	=&\sum_q p(q) [I(V,V_{2};Y_2|U,Q=q)-I(V,V_{2};Z|U,Q=q)]&\nonumber\\
	=&p(\alpha) [I(V_\alpha,V_{2\alpha};Y^*_2|U_\alpha)-I(V_\alpha,V_{2\alpha};Z^*|U_\alpha)]&\nonumber\\&+p(\beta) [I(V_\beta,V_{2\beta};Y^*_2|U_\beta)-I(V_\beta,V_{2\beta};Z^*|U_\beta)]&\nonumber\\
	=&\gamma [I(V^*,V^*_{2};Y^*_2|U^*)-I(V^*,V^*_{2};Z^*|U^*)]&\nonumber\\&+(1-\gamma) [I(U^*;Y^*_2|U^*)-I(U^*;Z^*|U^*)]&\nonumber\\
	=&\gamma [I(V^*,V^*_{2};Y^*_2|U^*)-I(V^*,V^*_{2};Z^*|U^*)]&
	\end{flalign}
	From (\ref{eq7}.i),\\
	\begin{flalign}\label{eq22}
	0<&I(V'_1;Y_1|V')-I(V'_1;Z|V')&\nonumber\\
	=&I(V_1,Q;Y_1|V,Q)-I(V_1,Q;Z|V,Q)&\nonumber\\
	=&I(V_1;Y_1|V,Q)-I(V_1;Z|V,Q)&\nonumber\\
	=&\sum_q p(q) [I(V_{1};Y_1|V,Q=q)-I(V_{1};Z|V,Q=q)]&\nonumber\\
	=&p(\alpha) [I(V_{1\alpha};Y^*_1|V_\alpha)-I(V_{1\alpha};Z^*|V_\alpha)]&\nonumber\\&+p(\beta) [I(V_{1\beta};Y^*_1|V_\beta)-I(V_{1\beta};Z^*|V_\beta)]&\nonumber\\
	=&\gamma [I(V^*_{1};Y^*_1|V^*)-I(V^*_{1};Z^*|V^*)]&\nonumber\\&+(1-\gamma) [I(V^*,V^*_{1};Y^*_1|U^*)-I(V^*,V^*_{1};Z^*|U^*)]&
	\end{flalign}
	\begin{flalign}\label{eq23}
	0<&I(V'_2;Y_2|V')-I(V'_2;Z|V')&\nonumber\\
	=&I(V_2,Q;Y_2|V,Q)-I(V_2,Q;Z|V,Q)&\nonumber\\
	=&I(V_2;Y_2|V,Q)-I(V_2;Z|V,Q)&\nonumber\\
	=&\sum_q p(q) [I(V_{2};Y_2|V,Q=q)-I(V_{2};Z|V,Q=q)]&\nonumber\\
	=&p(\alpha) [I(V_{2\alpha};Y^*_2|V_\alpha)-I(V_{2\alpha};Z^*|V_\alpha)]&\nonumber\\&+p(\beta) [I(V_{2\beta};Y^*_2|V_\beta)-I(V_{2\beta};Z^*|V_\beta)]&\nonumber\\
	=&\gamma [I(V^*_{2};Y^*_1|V^*)-I(V^*_{2};Z^*|V^*)]&\nonumber\\&+(1-\gamma) [I(U^*;Y^*_1|U^*)-I(U^*;Z^*|U^*)]&\nonumber\\
	=&\gamma [I(V^*_{2};Y^*_1|V^*)-I(V^*_{2};Z^*|V^*)]&
	\end{flalign}
	For any $p(u^*)p(v^*|u^*)p(v^*_1,v^*_2|v^*)p(x^*|v^*_1,v^*_2) p(y^*_1,y^*_2,z^*|x^*)$ and any $\epsilon>0$, the points $(\varepsilon,\varepsilon)$ and $(I(V^*,V^*_1;Y^*_1|U^*)-I(V^*,V^*_1;Z^*|U^*)-\varepsilon,\varepsilon)$ satisfy the constraints (\ref{eq13})--(\ref{eq23}) for some sufficiently small $\gamma>0$. Likewise, in order to show that the point $(\varepsilon,D-\varepsilon)$ lies in $\mathcal{R}_{0}^{\text{new2}}$, we can present a similar argument by replacing $V_{2\beta}=V_\beta=U_\beta=U^*$ and $V_{1\beta}=(V^*,V^*_1)$ with $V_{1\beta}=V_\beta=U_\beta=U^*$ and $V_{2\beta}=(V^*,V^*_2)$. This proves that $\mathcal{R}_{0}^{\text{new2}}$ also goes arbitrarily close to the $R_2$ axis.

	Prior to proving condition 3, we note that $\mathcal{R}^{\text{new1}}$ is convex following our argument for condition 1. Recall in Theorem~\ref{theorem7}, we have $\text{co}(\mathcal{R}^{\text{new2}})=\mathcal{R}^{\text{new1}}$. We also proved condition 2 which states that $\mathcal{R}_{0}^{\text{new2}}$ contains all the interior points of $\mathcal{R}^{\text{new1}}$. Referring to Fig.~\ref{fig:8a}, $\mathcal{R}^{\text{new1}}\setminus \mathcal{R}_{0}^{\text{new2}}$ gives us the boundary points of $\mathcal{R}^{\text{new1}}$ along the $R_1$ and $R_2$ axes, i.e., all axes points from $(0,0)$ to $s_2$ as well as from $(0,0)$ to $s_4$. In the proof of Theorem~\ref{theorem7}, we have also shown that $(0,0),s_2\in\mathcal{R}_{2}^{\text{new2}}$. Due to the convexity of $\mathcal{R}_{2}^{\text{new2}}$, the line segment joining $(0,0)$ and $s_2$ falls in ${R}_{2}^{\text{new2}}$. Likewise, the line segment joining $(0,0)$ and $s_4$ falls in $\mathcal{R}_{1}^{\text{new2}}$ due to the same argument. Hence, we obtain condition 3. 
	
	Upon proving the three conditions, we have $\mathcal{R}^{\text{new2}}=\mathcal{R}_{0}^{\text{new2}}\cup \mathcal{R}_{1}^{\text{new2}}\cup \mathcal{R}_{2}^{\text{new2}}=\mathcal{R}^{\text{new1}}=\text{co}(\mathcal{R}^{\text{new2}})$ after combining Theorem~\ref{theorem6} and Theorem~\ref{theorem7}.
\end{IEEEproof}

\begin{corollary}\label{corollary 3}
	cl($\mathcal{R}_{0}^{\text{new2}}$)=cl($\mathcal{R}^{\text{old}}$).
\end{corollary}

\begin{IEEEproof}[Proof of Corollary~\ref{corollary 3}]
	Corollary~\ref{corollary 3} follows from the fact that $\mathcal{R}_{0}^{\text{new2}}$ contains all the interior points of $\mathcal{R}^{\text{new1}}$ as stated in condition 2 for the proof of Theorem~\ref{theorem8}.
\end{IEEEproof}

\section{Discussions and Conclusion}

\begin{figure}[t]
	\centering
	\includegraphics[width=0.35\textwidth]{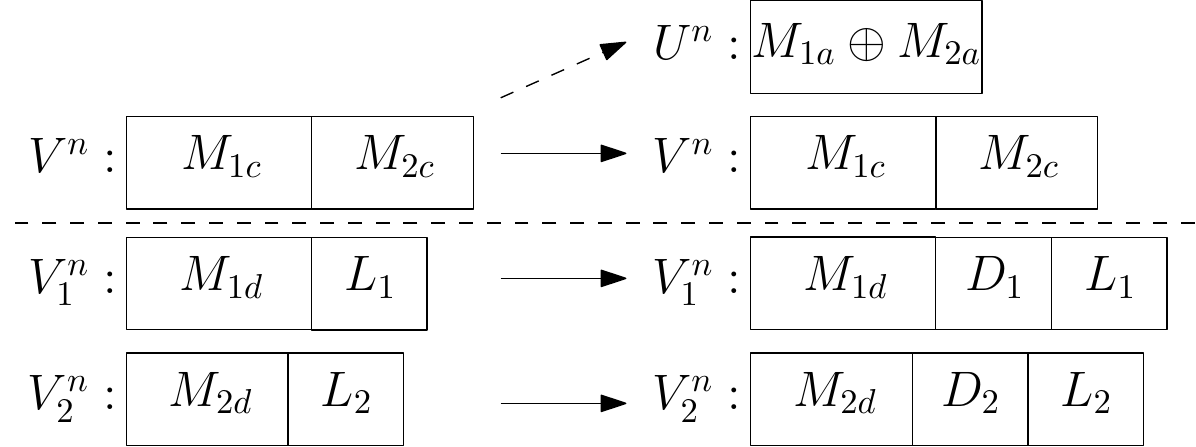}
	\caption{A secrecy technique integration strategy: Superposition-Marton coding scheme without secrecy (left); superposition-Marton coding scheme with secrecy (right).}
	\label{fig:9}
\end{figure}

For the case of the two-receiver discrete memoryless broadcast channel with complementary receiver side information where there is a passive eavesdropper, our proposed scheme simplification shows that the usage of Carleial-Hellman secrecy coding \cite{cref3} for the common satellite codewords achieves the same result as Wyner secrecy coding \cite{cref4}. The usage of Carleial-Hellman secrecy coding \cite{cref3} drops the usage of one-time pad signal for randomness, effectively reducing the number of message splits. The reduction in number of message splits simplifies the rate region computation process as the number of messages grows with the number of receivers in multi-receiver broadcast channels. Not least, no additional random component is needed to ensure individual secrecy of message segments in the common satellite codeword since each of the message segments is capable of fulfilling this requirement. The generation of random variables from a uniform distribution maximizes the difficulty of eavesdroppers in tapping the information. However, perfect uniformity is often difficult to achieve and slight information leakage is still probable \cite{cref10}. Hence, this reduction in the number of random components helps lower the probability of information leakage due to non-uniform distributions and results in a more secured secrecy coding scheme. 

The proposed simplified coding scheme has a simple construction which allows us to identify an intuitive strategy on applying secrecy techniques to error-correcting coding schemes under individual secrecy constraints. Referring to the left figure in Fig.~\ref{fig:9}, when secrecy is not required, the best achievability scheme for broadcast channels in up-to-date literature would be the superposition-Marton coding scheme. In this scheme, the $V^n$ codeword carries common messages which will be decoded by all receivers. Meanwhile, the remaining $V_1^n$ and $V_2^n$ codewords will only carry private messages intended to be decoded at the respective receivers. 

Now, when individual secrecy sets in, we notice that secrecy techniques can be integrated into the superposition-Marton coding scheme without changing its structural framework. As shown in the right figure in Fig.~\ref{fig:9}, the private message codewords $V_1^n$ and $V_2^n$ can be secured with Wyner secrecy coding \cite{cref4} through the introduction of random components $D_1$ and $D_2$. Meanwhile, the common message codeword $V^n$ can also be protected from eavesdroppers by implementing Carleial-Hellman secrecy coding \cite{cref3} and have the message segments $M_{1c}$ and $M_{2c}$ act as random components to protect each another. It is also interesting to see that the availability of receiver side information allows us to integrate the one-time pad signal \cite{cref8}, which is a network layer solution, to our physical layer coding scheme. By forming an additional common message codeword $U^n$ which carries the one-time pad signal, we are able to cover extreme cases in which the channel to the eavesdropper is strictly stronger than the channel to the legitimate receiver(s). Lastly, it is important to note that although our discussion revolves around the case of two-receiver broadcast channels, this secrecy technique integration strategy should hold for broadcast channels with more than two receivers if individual secrecy is required and the superposition-Marton coding scheme is still applied.

\enlargethispage{-0.8cm} 

\bibliographystyle{IEEEtran}
\bibliography{citations}

	







\end{document}